\documentclass[sigconf, nonacm]{acmart}

\usepackage{amsfonts}
\usepackage{verbatim}
\usepackage{graphicx,subfigure,multirow}
\usepackage{epstopdf}
\usepackage{comment}
\usepackage{array}
\usepackage{mathtools}
\usepackage{amsmath}
\usepackage{color}
\usepackage{multirow}
\usepackage{url}
\usepackage{eurosym}
\usepackage{booktabs} 
\usepackage{xr}

\makeatletter
\newif\if@restonecol
\makeatother

\usepackage[ruled,vlined,linesnumbered]{algorithm2e}
\usepackage{algorithmic}

\newcommand{\hide}[1]{} 
\newcommand{\vpara}[1]{\vspace{0.05in}\noindent\textbf{#1 }}

\DeclareMathOperator*{\argmin}{arg\,min}

\newcommand{\xw}[1]{\noindent{\textcolor{blue}{**XW:~#1**}}}

\newcommand{\eg}{{\sl e.g.}}
\newcommand{\ie}{{\sl i.e.}}

\newcommand{\etal}{{\sl et al.}}

\newcommand{\dataset}{{CWS}}

\def\bb{{\mathbf{b}}}

\def\bx{{\mathbf{x}}}
\def\by{{\mathbf{y}}}
\def\bz{{\mathbf{z}}}
\def\bX{{\mathbf{X}}}

\def\bq{{\mathbf{q}}}

\def\bW{{\mathbf{W}}}

\def\balpha{{\boldsymbol{\alpha}}}

\def\bpi{{\boldsymbol{\pi}}}
\def\bPi{{\boldsymbol{\Pi}}}
\def\cD{{\mathcal{D}}}

\def\cL{{\mathcal{L}}}

\def\RR{{\mathbb{R}}}

\def\II{{\mathbb{I}}}

\begin{document}

\title{Interpretable Learning-to-Rank with Generalized Additive Models}

\author{Honglei Zhuang, Xuanhui Wang, Michael Bendersky, Alexander Grushetsky, Yonghui Wu, 
Petr Mitrichev, Ethan Sterling, Nathan Bell, Walker Ravina, Hai Qian}
\affiliation{
  \institution{Google}
}



\renewcommand{\shortauthors}{Zhuang et al.}

\sloppy

\begin{abstract}

Interpretability of learning-to-rank models is a crucial yet relatively under-examined research area.
Recent progress on interpretable ranking models largely focuses on generating \emph{post-hoc} explanations for existing black-box ranking models, 
whereas the alternative option of building an \emph{intrinsically interpretable} ranking model with transparent and self-explainable structure remains unexplored.
Developing fully-understandable ranking models is necessary in some scenarios (e.g., due to legal or policy constraints) where post-hoc methods cannot provide sufficiently accurate explanations~\cite{rudin:nature2019}.

In this paper, we lay the groundwork for intrinsically interpretable learning-to-rank by introducing generalized additive models (GAMs) into ranking tasks.  
Generalized additive models (GAMs) are intrinsically interpretable machine learning models and have been extensively studied on regression and classification tasks.
We study how to extend GAMs into ranking models which can handle both item-level and list-level features and propose a novel formulation of \emph{ranking GAMs}. 
To instantiate ranking GAMs, we employ neural networks instead of traditional splines or regression trees.
We also show that our neural ranking GAMs can be distilled into a set of simple and compact piece-wise linear functions that are much more efficient to evaluate with little accuracy loss. 
We conduct experiments on three data sets and show that our proposed neural ranking GAMs can achieve significantly better performance than other traditional GAM baselines while maintaining similar interpretability.

\end{abstract}


\maketitle

\section{Introduction}
\label{sec:intro}

\begin{figure}[t]
\centering
\includegraphics[width=1.0\linewidth]{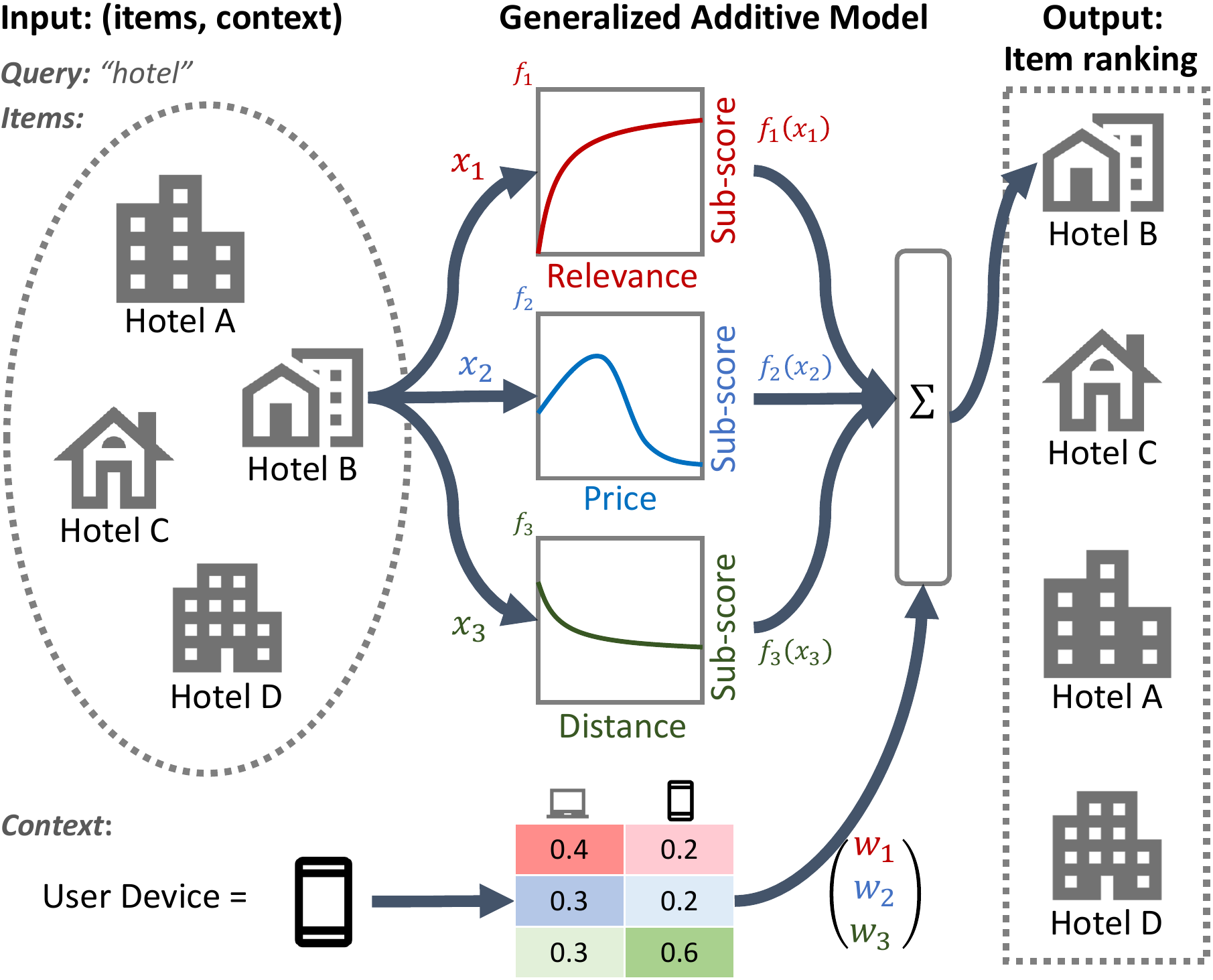}
\caption{
An example of a ranking GAM for local search.
For each input feature $x_j$ (\eg~price, distance), a sub-model produces a sub-score $f_j(x_j)$.  
Context features (\eg~user device type) can be utilized to derive importance weights of sub-models.
The ranking score of each item is a weighted sum of sub-scores.
The output is a ranked list of items sorted by their ranking scores.
}
\label{fig:example}
\end{figure}

Learning-to-rank (LTR)~\cite{liu2009ltr} has been extensively studied~\cite{MART:friedman2001greedy,LightGBM:ke2017nips,burges2010ranknet,pasumarthi2019tf,burges2005learning,burges2007learning} with broad applications in various fields~\cite{macdonald2013whens,karatzoglou2013learning}.
While many studies focus on building more accurate ranking models, few pay attention to the model interpretability.
The lack of interpretability can lead to many issues in practice, 
such as difficult troubleshooting, opacity to potential social bias~\cite{kulshrestha2019search}, vulnerability to feature corruption, and high maintenance costs~\cite{sculley2015hidden}.
The latest SWIRL workshop report~\cite{culpepper:sigir2018} lists model transparency as one of the key issues that the IR community should focus on in the following years.  
Hence, there is an emerging need to develop more transparent and interpretable learning-to-rank models.

There are roughly two categories of techniques to achieve model interpretability~\cite{du:cacm2019}:
one aims to build \emph{intrinsically interpretable} models where the model structure per se is understandable; the other aims to generate \emph{post-hoc} explanations of an existing black-box model.
In the specific application of learning-to-rank, there are a few recent studies~\cite{singh2019exs,singh2018posthoc,verma2019lirme,rennings2019axiomatic} focusing on generating \emph{post-hoc} explanations for existing models, whereas \emph{intrinsically interpretable} learning-to-rank models remain relatively unexplored\footnote{Tree-based models~\cite{MART:friedman2001greedy,LightGBM:ke2017nips,burges2010ranknet} are often considered more interpretable than neural models.  However, the state-of-the-art implementations of tree-based models, like LambdaMART, often involve thousands of deep and wide trees, tree ensembles and  bagging~\cite{burges2011learning}, making them virtually impossible to interpret by humans.}.

Although the post-hoc explanations are valuable and insightful for some applications,
they can also suffer from being inaccurate about actual model behaviors, insufficient to understand model details or unable to incorporate information outside of the database, as pointed out in a recent study~\cite{rudin:nature2019}. Hence, it is necessary to develop an intrinsically interpretable LTR model as an alternative for scenarios where ranking models need to be fully understandable by a human observer.

For example, consider the cases where a ranking system is involved in determining bail or parole, deployment of police in city neighborhoods, loan eligibility assessment, advertisement targeting, or guiding medical treatement decisions -- all of which are highly plausible scenarios today~\cite{alex2018frontiers,Obermeyer447,Mehrabi2019ASO}. In such high-stakes decision-making scenarios, researchers are morally (and often legally) obligated to develop and deploy interpretable models. The contribution of \emph{each individual feature} in these models should be examinable and understandable, to ensure transparency, accountability and fairness of the outcomes.

Generalized additive models (GAMs)~\cite{hastie1986generalized} provide a promising option to build such intrinsically interpretable models and have been used in many domains like finance~\cite{berg2007bankruptcy} and health~\cite{caruana2015intelligible,zhang2019axiomatic}. 
The output of a GAM is the sum of the outputs of multiple sub-models, where each sub-model only takes an individual feature as input.
A GAM has a higher model capacity than a linear model as sub-models can be more flexible than linear functions, while staying intelligible comparing with a black-box model. 
Each sub-model of a GAM precisely reflects the contribution of each input feature to the final output results.
Previous studies~\cite{lou2012intelligible,zhang2019axiomatic} primarily applied GAMs to regression or classification.
However, to the best of our knowledge, how to build GAMs for \emph{learning-to-rank} tasks has not been well-studied.

A learning-to-rank task has different input and objective from classification or regression.
The goal is to rank a list of items given some context. 
Hence, there would be both item-level features and list-level context features available. 
In Figure~\ref{fig:example}, we show an example of local search. 
Based on a user query ``hotel'', a set of item-level features $x_j$'s are computed (\eg~relevance score and distance) for each item (hotel). 
In addition, there could be list-level context features (\eg~user device type). 
A ranking GAM should not only inherit the intelligibility of GAMs, but can also take the context features into account. 
As presented in Figure~\ref{fig:example}, the ranking GAM consists of several isolated sub-models, where each sub-model only takes a single item-level feature $x_j$ as input and outputs a corresponding sub-score $f_j(x_j)$. 
In addition, 
the context features are used to derive importance weights for item-level sub-scores before they are summed up as the final ranking score.
This ensures that not only the contribution of each item-level feature is interpretable, but also the contributions of the context features, which is crucial in the high-stakes decision-making scenarios.

There are multiple challenges to adapt existing GAMs for ranking tasks. As we will show later, directly projecting the context features from list-level to item-level is not effective for ranking tasks. How to leverage context features (\eg~user device type) in a ranking GAM is not clear.
Moreover, traditional GAMs are only trained with loss functions for regression or classification tasks.  It is unclear how effective such models can be trained with ranking losses.


To overcome these challenges, we propose a novel neural GAM model for learning-to-rank tasks. 
In such a model, we employ standalone neural networks to instantiate sub-models for each individual feature.
The flexibility of neural network framework enables us to implement the ranking GAM structure which incorporates list-level context features and optimize the model with ranking losses.
The model can also seamlessly handle both numerical and non-numerical features.

In summary, we make the following contributions in this paper:
\begin{itemize}
    \item We formally define the structure of ranking GAMs in both the context-absent and context-present scenarios. 
    \item We propose to instantiate ranking GAMs by neural networks using a novel neural architecture that can leverage context features effectively and can be trained with ranking losses.
    \item We develop a novel technique to distill each sub-model of a ranking GAM into a simple piece-wise linear function.  
    \item We evaluate our proposed neural ranking GAMs on three learning-to-rank data sets and show the effectiveness.  
\end{itemize}

\section{Related Work}
\label{sec:related}


\vpara{Interpretable learning-to-rank.}
Model interpretability~\cite{zhang2019tutorial} becomes a popular research topic in recent years~\cite{lundberg2017unified,sundararajan2017axiomatic,shrikumar2017learning,ribeiro2016should,wu2018beyond,chen2018neural}. 
Current studies on interpretable learning-to-rank mainly focus on generating interpretation for an existing black-box model.  Singh~\etal~\cite{singh2019exs} describe a system to quantify and visualize the effect of each term on a certain decision of a given ranking model, such as why a document is ranked higher than another.  
They also attempt to distill a given ranking model with a subset of interpretable features in~\cite{singh2018posthoc}.
Verma~\etal~\cite{verma2019lirme} also study a similar problem to quantify the contribution of terms in a specific document ranking generated by a given ranking model.  
Rennings~\etal~\cite{rennings2019axiomatic} propose a diagnostic data set based on a set of axioms to identify potential shortcomings in an existing IR system. However, there are few studies specifically designed to build ``intrinsically'' interpretable ranking models where each component of the model is understandable.

\vpara{Generalized additive model.}
Proposed by Hastie~\etal~\cite{hastie1986generalized}, generalized additive models (GAMs) provide a class of models which are more flexible than linear models, yet remain more interpretable than complicated models such as deep neural networks. 
Binder~\etal~\cite{binder2008comparison} provide a thorough study on fitting GAMs with regression splines.
Lou~\etal~\cite{lou2012intelligible} also compare GAMs with different function instantiations.
They extend the comparing methods to include regression trees and conduct experiments on both classification and regression tasks.  
Their studies shows that shallow bagged trees perform better comparing to other functions such as splines.
Zhang~\etal~\cite{zhang2019axiomatic} further studies to train GAMs specifically for multi-class classification tasks. 

To obtain more accurate models, higher-order feature interactions were introduced to the additive models.
Lou~\etal~\cite{lou2013accurate} propose an efficient algorithm to identify feature pairs with interactions.  
A series of studies by Hooker~\cite{hooker2004discovering,hooker2007generalized} also aim to discover additive structure with higher-order feature interactions.  
Meanwhile, some other studies tackle the problem in high-dimensional data where the number of features is huge.  
Others aim to reduce the number of features utilized in the model~\cite{ravikumar2009sparse,lin2006component,meier2009high,petersen2019data} by imposing sparsity penalty.  
Lou~\etal~\cite{lou2016sparse} introduce to partially replace the non-linear function of some features by linear function to reduce the model complexity.
They provide an algorithm to automatically select which features should use linear functions and which should use non-linear functions. 
Chouldechova~\etal~\cite{chouldechova2015generalized} also explore similar ideas. However, to the best of our knowledge, none of these studies focus on applying GAMs to learning-to-rank problems.



\section{Problem Formulation}
\label{sec:prelim}

In this section, we start by defining notations for a learning-to-rank problem.  
Then we formalize generalized additive model (GAM) and propose a novel formulation of ranking GAM.  


\subsection{Learning to Rank}
In a ranking problem, we denote a data set with $N$ lists as $\cD=\{(\bq, \bX, \by)\}_{1}^{N}$.
For a specific list $(\bq, \bX, \by) \in \cD$,
$\bq=(q_1, \cdots, q_m)$ is a context feature vector consisting of $m$ list-level contextual signals (\eg~query features in search tasks, user information in recommendation tasks);
$\bX=\{\bx_{i}\}_{i=1}^{l}$ is a set of $l$ data items, each represented by a feature vector $\bx_{i}$;
$\by=\{y_{i}\}_{i=1}^{l}$ is a set of relevance labels of corresponding data items where a higher $y_{i} \in \RR$ indicates that the item $\bx_{i}$ is more relevant.
Let $\bPi_{l}$ denote the set of all permutations of $l$ data items. The optimal ranking $\bpi^* \in \bPi_{l}$ can always be obtained by ranking items $\bx_{i}$'s according to their relevance labels $y_{i}$'s from highest to lowest.


The typical learning-to-rank setting aims to learn a ranker $\varphi$ from a training data set $\cD_L$ with given relevance labels, such that for any list $(\bq, \bX)$, the inferred ranking $\hat{\bpi} = \varphi (\bq, \bX)$ can be as close to the optimal ground-truth ranking $\bpi^*$ as possible. 

Specifically, in this work, we infer the ranking by scoring each item individually and sorting them based on their scores.
For each list $(\bq, \bX)$ with given context features and item features, 
we aim to learn a scoring function $F$ which takes both the context features $\bq$ and an individual item's features $\bx_{i}$ as input, and output a ranking score $\hat{y}_{i} \in \RR$:
\begin{align}
    \hat{y}_{i} = F(\bq, \bx_{i}) \nonumber
\end{align}
The final predicted ranking $\hat{\bpi}$ will simply be generated by ranking all the items in $\bX$ based on their inferred ranking scores $\hat{y}_{i}$'s.


It is worth noting that there is not yet a common understanding on a formal definitions of interpretability in learning-to-rank.
In this paper, we confine our discussion within the interpretability offered by generalized additive models (GAMs).
We will briefly review the formal definition of GAMs, and then propose a novel ranking GAM definition with similar interpretability.

\subsection{GAM}
A generalized additive model (GAM)~\cite{hastie1986generalized,lou2012intelligible} learns a function for each individual input feature respectively.  
Previous studies typically focus on applying generalized additive models on classification or regression tasks with numeric features.  
Formally, we denote a data set as $\cD=\{(\bx_i, y_i)\}_{i=1}^{N}$ where each $\bx_i = (x_{i1}, \cdots, x_{in})$ is a feature vector containing $n$ features and $y_i$ is the target.  
For a regression task, $y_i \in \RR$ is a real value, while for a binary classification task $y_i \in \{0, 1\}$ is a binary value.  
A generalized additive model takes a feature vector $\bx_i$ and outputs $\hat{y}_i$ with the following structure:
\begin{align}
    g(\hat{y}_i) = f_1(x_{i1}) + f_2(x_{i2}) + \cdots + f_n(x_{in}) \label{eq:gam}
\end{align}
Each $f_j(\cdot)$ is a function to be learned for the $j$-th feature.
The function can be instantiated by different classes of functions, such as linear functions, splines or trees/ensemble of trees~\cite{lou2012intelligible}. 
$g(\cdot)$ is a link function that links the sum of all $f_j(x_{ij})$'s to the model output. 
For example, $g(\cdot)$ could be an identity function for regression tasks, while for a binary classification task its inverse function $g^{-1}(\cdot)$ could be a logistic function $g^{-1}(u) = \frac{1}{1 + \exp(-u)}$.

The model does not involve any interactions between features, which could lead to compromise on performance.
However, the simple structure also introduces many appealing benefits in practices.
Since each $f_j(\cdot)$ is essentially a univariate function, the relationship between a feature's value and the final response can be accurately quantified and visualized by plotting $f_j(\cdot)$.
This transparency can greatly reduce the cost for model troubleshooting and maintenance.

\hide{
One of the most essential advantages of the additive model structure is the interpretability.  
Since each $f_j(\cdot)$ is essentially a univariate function, one can quantify and visualize how a single feature's value contributes to the final predicted response by plotting $f_j(\cdot)$. 
The additive structure also makes the model easier to be edited by adjusting the shape of $f_j(\cdot)$.   
For example, in some scenarios, the contribution of a particular feature cannot surpass a certain threshold. This can be done by clipping the output of $f_j(\cdot)$ in the model.  
}

\subsection{Ranking GAM}
Traditional GAMs are designed for classification and regression problems. There are no systematic studies to develop GAMs for ranking problems.  
In this subsection, we define ranking GAMs, which are designed to tackle the special structure of ranking problems while maintaining intelligibility of traditional GAMs. 

\vpara{Context-absent ranking.}
We start our discussion by the context-absent ranking scenario, where the list-level context features $\bq$ are not available.  In this scenario, a ranking GAM essentially applies a traditional GAM for regression on each item $\bx_i$ in the list $\bX$ as a scoring function.
Given each item's representation $\bx_{i}=(x_{i1}, \cdots, x_{in})$, the ranking score $\hat{y}_i$ can be derived by:
\begin{align}
    \hat{y}_i = F(\bx_{i}) = f_1(x_{i1}) + f_2(x_{i2}) + \cdots + f_n(x_{in}) \label{eq:garm}
\end{align}

Note that the item-level features can be context-dependent. For example, the BM25 scores are item-level features but depend on both queries (contexts) and documents (items).

However, not all context features can be effectively projected to item-level features, e.g., the time in a day when a query is sent in search tasks (like 9 a.m. or 9 p.m.).  
In this case, the context-absent ranking formalization is not compatible with such a context feature.

\vpara{Context-present ranking.}
Now we discuss the definition of our ranking GAM in the context-present scenario where list-level context features $\bq=(q_1, \cdots, q_m)$ are available.  
A straightforward solution would be to project context features $\bq$ as item-level features and apply a traditional GAM as an item scoring function like in the context-absent setting:
\begin{align}
    \hat{y}_i = F(\bq, \bx_{i}) &= f_1(x_{i1}) + f_2(x_{i2}) + \cdots + f_n(x_{in}) \nonumber \\
    &  + f_{n+1}(q_1) + f_{n+2}(q_2) + \cdots + f_{n+m}(q_m) \label{eq:cngarm_bad_example}
\end{align}
Unfortunately, this model structure cannot fully leverage the signals from context features in learning-to-rank scenarios.
On one hand, most ranking losses used for training only involve the differences of predicted ranking scores $(\hat{y}_{i} - \hat{y}_{i'})$ between items.  
Obviously, the sub-model terms of context features $f_{n+k}(q_k)$ will be canceled out as all items within the same list share the same context feature values.  
On the other hand, most ranking metrics only care about the order of items $\hat{\bpi}$, which is not related to context feature sub-model terms $f_{n+k}(q_k)$ either. 

In order to leverage context features for a GAM model, we propose the following ranking GAM model that 
utilizes context features to derive importance weights when combining item-level features.
Specifically, the predicted ranking score of each item is:
\begin{align}
    \hat{y}_i = F(\bq, \bx_{i}) &= \sum_{j=1}^{n} \sum_{k=1}^{m} w_{j, k} (q_k) f_{j} (x_{ij})  \label{eq:cngarm}
\end{align}
where both $f_{j} (\cdot)$ and $w_{j, k}(\cdot)$ are arbitrary univariate functions to be learned.  
It is worth noting that when the context features $\bq$ are fixed, the predicted ranking score can still be decomposed as sum of functions of each item-level feature:
\begin{align}
    \hat{y}_i = F(\bq, \bx_{i}) &= \sum_{j=1}^{n} \Big(  w_j(\bq) f_{j} (x_{ij}) \Big)  \label{eq:cngarm_decompose_f}
\end{align}
where $w_j(\bq) = \sum_{k=1}^{m} w_{j, k} (q_k)$ can be interpreted as the importance weight of the $j$-th item feature derived from all the context features. For example, in a search task, item features like distance might be more important if users are searching for hotels, while the content relevance might be more important if users are searching for a convention center. And notice that the weight function $w_j(\bq)$ is also an additive model with regard to each context feature.

\section{Neural Ranking GAM}
\label{sec:model}

In this section, we propose our method that instantiates ranking GAM based on neural networks.


\subsection{Context-Absent Neural Ranking GAM}
\label{subsec:ngam}

We start with the context-absent scenario, where a ranking data set can be represented as $\cD=\{(\bX, \by)\}_{1}^{N}$ as the context features $\bq$ are not available.
We will build a scoring function that predicts the ranking score $\hat{y}_{i}$ for each data item $\bx_{i}$ in each list $\bX$.

For simplicity, we focus on a single data item in a list and omit the subscripts.  
A data item can be represented as $\bx = \{x_1, x_2, \cdots, x_n\}$ where $x_j$ represents the $j$-th feature of the data item $\bx$.


We build a standalone neural network for each feature $x_i$ which outputs a single ``sub-score'' $s_i \in \RR$.
This leads to $n$ separate neural networks in total.
In principle, users have the flexibility to construct any neural network structure with legitimate input and output.
One can even build a different network structure for each feature according to its characteristics.

\begin{figure}[t]
\centering
\includegraphics[width=0.55\linewidth]{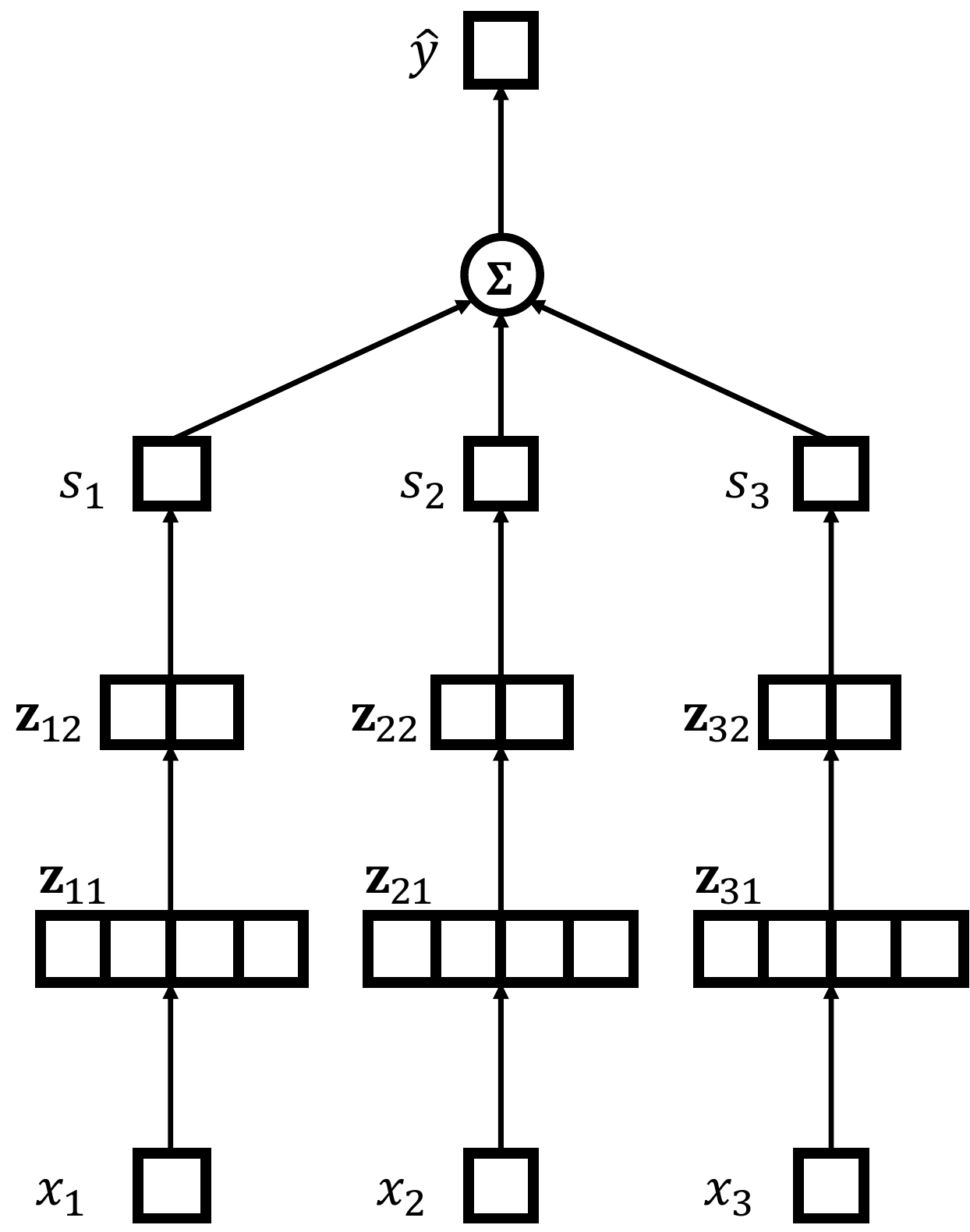}
\caption{
A graphical illustration of a context-absent neural ranking GAM.
}
\vspace{-0.2in}
\label{fig:ngam}
\end{figure}

In our practice, we simply adopt a feed-forward network structure with $L$ hidden layers for each feature as shown in Figure~\ref{fig:ngam}.
Specifically, for each item feature $x_j$, we can have 
\begin{align}
    \bz_{j1} &= \sigma(\bW_{j1} x_j + \bb_{j1}) \nonumber \\
    \bz_{j2} &= \sigma(\bW_{j2} \bz_{j1} + \bb_{j2}) \nonumber \\
    \ldots \nonumber \\
    \bz_{jH} &= \sigma(\bW_{jH} \bz_{j(H-1)} + \bb_{jH}) \nonumber 
\end{align}
where $\bz_{h}$ is the output of the $h$-th hidden layer; 
$\bW_{jh}$ and $\bb_{jh}$ are weight matrix and bias vector of the $h$-th hidden layer to be trained; 
$\sigma(\cdot)$ is the non-linear activation function. 
We choose the Rectifier (ReLU)~\cite{nair2010rectified} as the activation function.  

The sub-score of feature $x_j$ can be obtained by feeding the output of the final hidden layer into a dense layer:
\begin{align}
    f_j(x_j) = s_j = \bW_{j} \bz_{jH} + b_{j}  \label{eq:subscore}
\end{align}

Based on the $n$ neural networks we build for all the $n$ features, we can obtain the final predicted score for item $\bx$ by simply taking the sum of all the sub-scores:
\begin{align}
    \hat{y} = F(\bx) = \sum_j f_j(x_j) = \sum_j s_j   \label{eq:gam_sum}
\end{align}
And the final predicted ranking of list $\hat{\bpi}$ can be obtained by simply ranking all the items $\bx$'s in the list according to their predicted ranking scores.




\subsection{Context-Present Neural Ranking GAM}
\label{subsec:cngam}

In this subsection, we describe the neural network instantiation of the context-present ranking GAMs where context features $\bq$ of each list are available.
In this design, the module for item features remains similar as the context-absent setting.
In addition, we build another neural additive module for context features $\bq$ which outputs an importance weights vector for item feature sub-models instead of a score.  

\begin{figure}[t]
\centering
\includegraphics[width=0.99\linewidth]{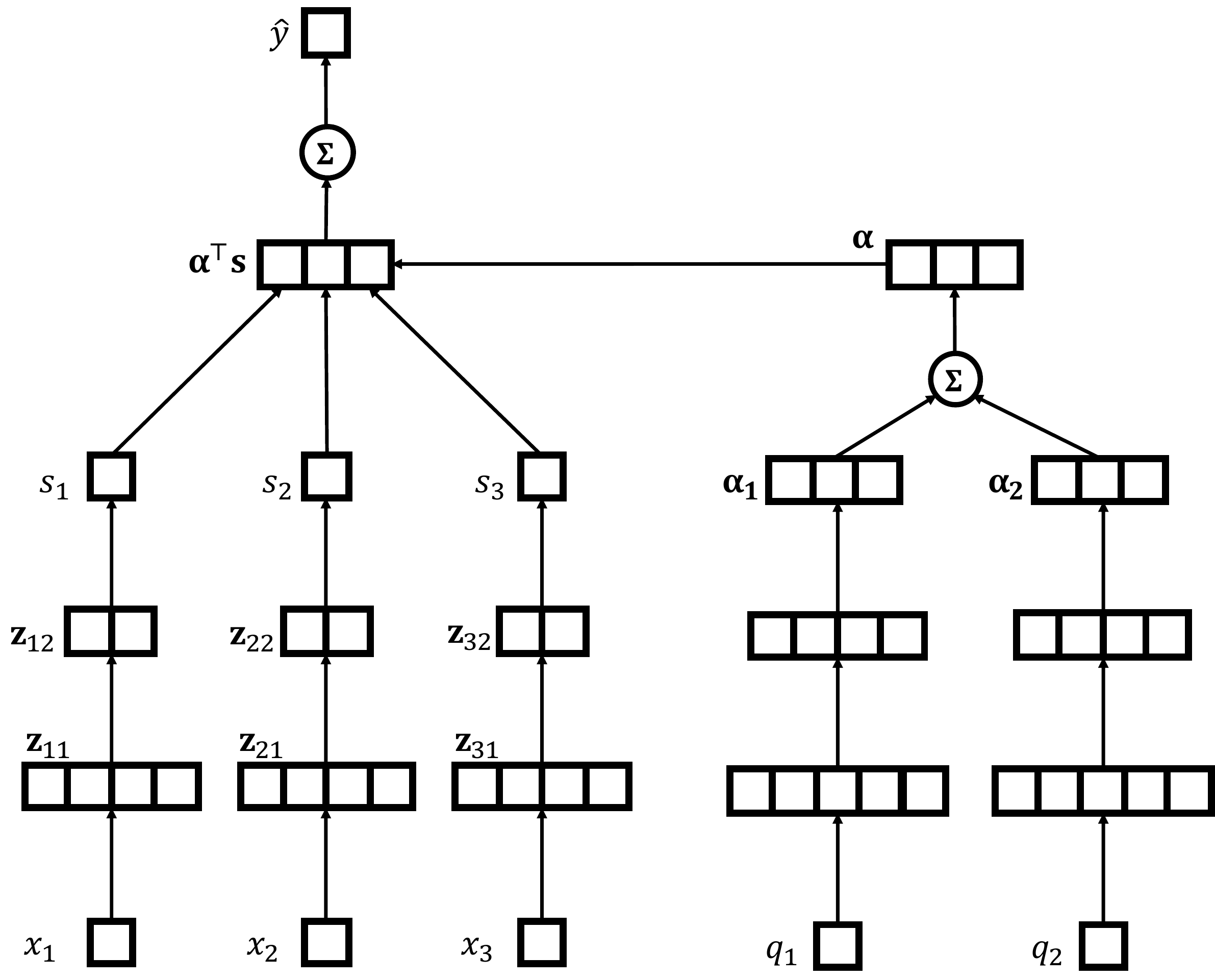}
\vspace{-0.05in}
\caption{
A graphical illustration of a context-present neural ranking GAM.
}
\vspace{-0.2in}
\label{fig:cngam_ranking}
\end{figure}

We again focus on the scoring of a single data item and temporarily omit the subscripts.
Specifically, an item in list $\bX$ is denoted as $\bx=(x_1, \cdots, x_n)$, and the context features are denoted as $\bq=(q_1, \cdots, q_m)$. 
The sub-score $f_j(x_j)$ for each item feature $x_j$ is instantiated in the same way (Equation~\eqref{eq:subscore}) as the context-absent model.
From each individual context feature $q_k$, we derive an $n$-dimensional weighting vector $\balpha_k$.
As shown in Figure~\ref{fig:cngam_ranking}, the weighting vector can be obtained by a $T$-layer feed-forward neural network structure, where
\begin{align}
    \bz_{k1} &= \sigma(\bW_{k1} q_k + \bb_{k1}) \nonumber \\
    \bz_{k2} &= \sigma(\bW_{k2} \bz_{k1} + \bb_{k2}) \nonumber \\
    \ldots \nonumber \\
    \bz_{kT} &= \sigma(\bW_{kT} \bz_{k(T-1)} + \bb_{kT}) \nonumber 
\end{align}
Similarly, $\bz_{kt}$'s are hidden layer outputs and $\bW_{kt}, \bb_{kt}$ are model parameters to be learned. 
Then we use a softmax layer on top of the final dense layer to derive $\balpha_k$.
\begin{align}
    \balpha_k = \text{softmax}(\bW_{k}\bz_{kT})
\end{align}
where the $j$-th dimension of the weighting vector $\balpha_k^{(j)}$ indicates the importance of the $j$-th item feature in $\bx$ considering the $k$-th context feature. 
Notice that $\balpha_k^{(j)}$ exactly corresponds to the value of function $w_{j,k}(q_k)$ in Equation~\eqref{eq:cngarm}.
The intuition of using a softmax layer is to prevent the derived importance weights to be negative or to be extremely large on some item features, which would substantially compromise the model interpretability.

We denote the overall weighting vector as $\balpha \in \RR^n$.  
We obtain $\balpha$ by taking the summation of weighting vectors $\balpha_k$ from all the $m$ context features in $\bq$:
\begin{align}
    \balpha = \sum_{k=1}^{m} \balpha_k
\end{align}
The final ranking score is generated by taking the weighted sum of sub-model $f_j(x_j)$'s as described in Equation~\eqref{eq:cngarm_decompose_f}, where the weight of the $j$-th item feature $w_j(\bq)$ equals to $\balpha^{(j)}$, \ie~the $j$-th dimension of the overall weighting vector.




\subsection{Remarks}
\vpara{Ranking losses.} Both neural ranking GAMs can be trained with any ranking loss functions.  
In this work, we train our model with approximate NDCG loss from~\cite{qin2010general,bruch2019SIGIR}.
It is worth noting that regression losses such as Mean Squared Error (MSE) $\cL(\by, \hat{\by})=||\by - \hat{\by}||^2$ can be also used as loss functions for ranking. 
We will compare traditional GAMs optimizing regression loss with the proposed ranking GAMs in our experiments.

\vpara{Non-numerical features.} Another advantage of adopting neural networks is the flexibility to handle non-numerical features such as categorical or textual features, since handling such features in traditional GAM instantiated by splines or trees is non-trivial.
Such features can be seamlessly incorporated into GAMs instantiated by neural networks by deriving embedding vectors.  
Notice that the introduction of embedding will not affect the final visualization of sub-models for non-numerical features.

\section{Sub-Model Distillation}
\label{sec:curve}

Model distillation~\cite{hinton2015distilling} is a widely-adopted technique to simplify deep neural networks.
The general idea is to train a smaller, simpler model by minimizing the loss between its output and that of a more larger, complex model. 
Given the special architecture of our proposed neural GAM model, 
we propose to distill each \emph{sub-model} individually.
The benefit of sub-model distillation is multi-fold:
the distilled sub-models can be faster to perform inference;
they are also smaller in size which is beneficial for on-device deployment;
they could potentially be more interpretable due to their simpler structure.
We focus on distilling sub-models of numerical features where we utilize piece-wise regression (also known as segmented regression)~\cite{segmented:regression}.


\subsection{Piece-wise Regression}
We are given a numerical feature $x$ and its sub-model $f(x)$ learned in a GAM.
The goal of piece-wise regression is to find a piece-wise linear function (PWL) that is as close to $f(x)$ as possible. 
A PWL function can be described by a set of $K$ knots, denoted as $S=\{(x_k, y_k)\}_{k=1}^K$ where $x_k$'s in the knots are ordered,~\ie~$x_k < x_{k+1}$.
The definition of PWL function based on the knots $S$ is:
\begin{equation*}
PWL_{S}(x) = 
    \begin{cases}
        y_1 & \text{if } x < x_1, \\
        \frac{y_{k+1} - y_k}{x_{k+1} - x_k} (x - x_k) + y_k & \text{if } x_k \leq x \leq x_{k+1}, \\
        y_K & \text{if } x > x_K.
    \end{cases}
\end{equation*}
$PWL_S(x)$ is a linear function between any two adjacent knots and becomes flat at two ends. 
Thus, the $K$ knots uniquely determine the PWL function.
For the practice of sub-model distillation, a small $K$ (\eg~around 3 to 5) is usually sufficient.

Formally, the goal of piece-wise regression is to find the set of optimal knots $S=\{(x_k, y_k)\}_{k=1}^K$ that minimize the empirical loss on a training data set $\cD = \{(x_i, f(x_i)\}_{i=1}^N$. 
In particular, we use the Mean Squared Error (MSE) as our loss function and have the following optimization problem:
\begin{equation} \label{eq:pwl_opt}
    S^* = \argmin_{S=\{(x_k, y_k)\}_{k=1}^K} \frac{1}{|\cD|} \sum_{\cD} \big\|f(x_i) - PWL_S(x_i)\big\|^2
\end{equation}
Note that knots $S=\{(x_k, y_k)\}_{k=1}^K$ are not necessarily a subset of the training data $\cD=\{(x_i, f(x_i)\}_{i=1}^N$.

Such a formulation can be solved a generic piece-wise regression package such as~\cite{segmented:package}. However, such a package is usually slow and not scalable to large data sets. In the next section, we propose a greedy algorithm for this problem.

\subsection{Fitting Algorithm}
The major challenge of our fitting algorithm is to determine the $x_k$'s for the $K$ knots: $\{x_k\}_{k=1}^K$. 
With $x_k$'s determined, the optimal $y_k$'s can be computed by a method similar to least squares, which we will not elaborate in this paper due to limited space.

Our method first generates a set of $x$'s as knot candidates $P$. Given all $x$'s in the training data $\cD=\{(x_i, f(x_i)\}$, we simply construct $P$ by a heuristic using percentile boundary of $0\%, 1\%, \cdots, 99\%, 100\%$ of $\cD$, resulting in at most $101$ elements in $P$.

With a given set of candidate knots $P$, our problem becomes a combinatorial optimization one, where one needs to find a subset $S \subset P$ with size $K$ that minimizes the loss in Equation~\eqref{eq:pwl_opt}. We denote the loss as $\cL(S)$.
A brute-force algorithm is to enumerate all the possible subset of $P$ with size $K$, but the computational cost is intractable.
Thus, we propose a greedy algorithm in Algorithm~\ref{alg:pwl_fitting}. 
The algorithm first constructs an initial set of knots in a greedy fashion and then refine it until convergence.
More specifically,
\begin{itemize}
    \item \textit{Knots Initialization}. 
    The algorithm greedily construct an initial set $S$ of size $K$ by adding $x$'s iteratively from $P$ (See Line~\ref{alg:init_start}-\ref{alg:init_end}).
    In each step, the $x^*$ that best minimizes the loss is added into the $S$.
    
    \item \textit{Knots Refinement}. 
    The algorithm iterates over each knot in $S$ and tries to assess whether it can be replaced by another candidate knot to further minimize the loss (See Line~\ref{alg:refine_start}-\ref{alg:refine_end}). 
    The procedure stops when there is no improvement after we loop over all $x_k \in S$ once. 
    
\end{itemize}

\begin{algorithm}[t]
\caption{Finding Knots for Piece-wise Regression} \label{alg:pwl_fitting}
\KwIn{Training data $\cD$, candidates $P$, output size $K$.}
\KwOut{Knots $\{(x_k, y_k)\}_{k=1}^K$.}
\begin{algorithmic}[1]

\STATE{Let $S \gets \{\min_{x \in P} x\}$}  \label{alg:init_start}
\FOR{$k \gets 2$ to $K$} 
    \STATE{$x^* \gets \argmin_{x \in P \setminus S} \cL(S \cup \{x\})$}
    \STATE{$S \gets S \cup \{x^*\}$}
\ENDFOR \label{alg:init_end}
\REPEAT   \label{alg:refine_start}  
    \FORALL{$x \in S$}
        \STATE{$x^* \gets \argmin_{x' \in P \setminus S} \cL(S \cup \{x'\} \setminus\{x\})$}
        \IF{$\cL(S \cup \{x^*\} \setminus\{x\})) < \cL(S)$} 
            \STATE{$S \gets S \cup \{x^*\} \setminus\{x\}$}
        \ENDIF

    \ENDFOR
\UNTIL{Convergence.} \label{alg:refine_end}
\RETURN{$\{(x_k, y_k) | x_k \in S\}$ after solving $y_k$'s for $S$.}
\end{algorithmic}
\end{algorithm}

Though the greedy algorithm may be stuck into local minimums in theory, we found that it outputs reasonably good PWL functions in practice as shown in our experiments. Since $K$ is usually very small, finding $y_k$ when $x_k$ are given can be done efficiently.


\section{Experiments}
\label{sec:exp}

In this section, we conduct experiments to answer the following two main research questions:
\begin{itemize}
    \item \textbf{RQ1}: How do our proposed neural ranking GAMs perform as compared to traditional GAMs on LTR problems?
    \item \textbf{RQ2}: How does the interpretability of neural ranking GAMs change comparing to traditional GAMs?
\end{itemize}
We first describe our experimental setup including data sets and comparing methods. Then we present the experimental results to answer the above research questions.

\begin{table*}[!th]
\centering
  \caption{
  \label{tab:data set}
    Basic statistics and properties of data sets.
  }
\vspace{-0.1in}
{
\small
\begin{tabular}{|c||c|c||c|c||c|c||c|c|}
\hline
\multirow{2}{*}{Data set}  & \multicolumn{2}{c||}{Train} & \multicolumn{2}{c||}{Validation}  & \multicolumn{2}{c||}{Test} & {Context} & {Non-numerical}\\ \cline{2-7}
 & Query & Document & Query & Document & Query & Document &  features? &  features? \\ \hline \hline
YAHOO  & 19,944 & 473,134 & 2,994 & 71,083 & 6,983 & 165,660 & No & No \\ \hline
WEB30K & 18,919 & 2,270,296 & 6,306 & 747,218 & 6,306 & 753,611 & No  & No\\ \hline
\dataset & 40,500 & 2,213,730 & 4,578 & 247,376 & 4,462 & 235,411 & Yes & Yes \\ \hline
\end{tabular}
}
\end{table*}

\subsection{Data Sets}
We use three data sets in our experiments. Two of them are public benchmark ones and they do not contain context features. The third one is a private data set that contains non-numerical context features. A summary of the data sets is shown in Table~\ref{tab:data set}.

\vpara{YAHOO.}
Yahoo! Learning to Rank Challenge data~\cite{chapelle2011yahoo} is a publicly available data set.
The original data set contains two sets for different purposes: Set1 and Set2. Set1 is the one commonly used for LTR evaluation. It consists of three partitions for training, validation and testing respectively. In this data set, each document is represented as up to 700 numerical features.
Each feature has been normalized into the range of $[0,1]$ by the inverse cumulative distribution. 
The meaning of each feature is not disclosed. There is no list-level context feature in this data set.  
Documents are labeled with 5-level relevance labels from 0 to 4.  


\vpara{WEB30K.}
WEB30K~\cite{qin2013letor} is a public learning-to-rank data set released by Microsoft.
We use Fold1 out of the total five folds in the original data.
Similarly to the YAHOO data set, Fold1 has the training, validation and testing partitions. In this data set, each document is represented by 136 numerical features.  
The meaning of each feature is disclosed and the range of each feature varies. Some query-dependent features are already captured by item-level features such as ``covered query term number''.
There is no additional list-level context feature available in this data set.  
Documents are also labeled with 5-level relevance labels from 0 to 4.


\vpara{Chrome Web Store (\dataset).}
\dataset{} is a private data set that we collected by sub-sampling from the Chrome Web Store logs. Each list in this data set corresponds to a visit to Chrome Web Store. The items in the list were the ones shown to a user and we collect the user actions like clicks or installations as our labels. For each item, we computed 15 numerical features. 
Moreover, each list contains 3 context features indicating users' region and language settings and the path of user visits and they are non-numerical. Similarly, 
we also partition the data set into three parts for training, validation and testing. 

\subsection{Experiment Setup}
\vpara{Comparing methods.}
The methods compared in our experiments are as follows:
\begin{itemize}
    \item \emph{Tree-based GAM (Tree GAM).}
    We train tree-based GAM with regression loss (mean squared error,~\ie, MSE).
    We use open-sourced software MLTK\footnote{\url{https://github.com/yinlou/mltk}} which implements GAM proposed by Lou~\etal~\cite{lou2012intelligible}.
    
    
    \item \emph{Tree-based Ranking GAM (Tree RankGAM).}
    We also try to train tree-based GAM with ranking loss.
    Since GAM has not been studied for ranking problems before, there is no existing implementations of GAM specifically designed for optimizing ranking loss.
    We use LambdaMART implemented by another open-sourced package\footnote{\url{https://github.com/Microsoft/LightGBM}} LightGBM~\cite{LightGBM:ke2017nips}, 
    and confine the depths of trees in the model to be $1$ so that the final model can be decomposed as Equation~\eqref{eq:garm}.
    The number of trees in LightGBM is set to $1,000$.

    
    \item \emph{Neural GAM (Neural GAM).}
    We use neural networks instead of trees to construct GAMs.  The model has similar structure with the context-absent neural ranking GAM, but is trained with regression loss (MSE) instead of ranking loss.
    
    \item \emph{Context-Absent Neural Ranking GAM (Neural RankGAM).}  The context-absent neural ranking GAM is presented in Section~\ref{subsec:ngam} which only takes item features. 
    We train the model with a ranking loss (approximate NDCG).
    

    \item \emph{Context-Present Neural Ranking GAM (Neural RankGAM$+$).}
    The context-present neural GAMs for ranking are presented in Section~\ref{subsec:cngam}. It can leverage context features to derive importance weights for item features.
    We also train the model with a ranking loss (approximate NDCG).
    
    
    
\end{itemize}

Our neural ranking GAMs are implemented based on the open-sourced TensorFlow Ranking library~\cite{pasumarthi2019tf}.

\vpara{Hyperparameters.}
For neural ranking GAMs, we tune the hyperparameters and choose the following in our experiments.
For YAHOO and WEB30K, we construct a neural network with $L=2$ layers for each item feature $x_i$ and the dimensions are $16$ and $8$ respectively.  
For \dataset~data set, we also build a neural network with $L=2$ layers for each item feature $x_i$ with dimensions as $64$ and $32$. 
Moreover, we build a neural network with $H=2$ layers for each context feature $q_i$ with dimensions as $128$ and $64$ for \dataset.
Since all context features in \dataset~data set are categorical, we use a $d$-dimensional  ($d=300$) embedding layer before the hidden layers for categorical features.  
We use AdaGrad~\cite{duchi2011adaptive} as the optimizer.

\vpara{Evaluation Metrics.}
Normalized Discounted Cumulative Gain (NDCG) is a standard metric for evaluating ranking quality with respect to items with graded relevance~\cite{jarvelin2002cumulated}.  
In our experiments, we utilize NDCG$_k$ as the evaluation metrics with different settings of $k$ (i.e., $1, 5, 10$) to evaluate the performance of each ranking method.

\subsection{Performance Comparison (RQ1)}
We first compare the performances of our proposed neural ranking GAMs to baselines.
Table~\ref{tab:performance} presents the overall performance comparison on all the tree data sets. 
We have the following observations.

The first message is the necessity of developing ranking GAMs specifically for LTR problems.
Compared with traditional GAMs which optimize regression loss, both Tree RankGAM and Neural RankGAM achieve substantial improvements. 
For example, on YAHOO, Neural RankGAM achieves NDCG$_5$ of 71.32\%, while the Neural GAM achieves only 69.62\%.  
On WEB30K, Neural RankGAM achieves NDCG$_5$ of 43.29\%, about +10\% higher than Neural GAM with regression loss. 
This confirms the importance to enable traditional GAMs to optimize ranking losses for LTR problems, as the performance can be extensively improved with exactly the same interpretability.


Moreover, on the \dataset~data set where context features are available, our proposed context-present neural ranking GAM (Neural RankGAM$+$) achieve significantly better performance than the context-absent version and Tree RankGAM. 
As one can observed, Neural RankGAM$+$ achieves +4.9\% improvement in terms of NDCG$_5$ comparing to the context-absent Neural RankGAM.  
The improvement shows that context-present Neural RankGAM$+$ can effectively leverage context features, which is a disadvantage of the traditional tree-based GAMs.  





\begin{table}[!t]
\centering
  \caption{
  \label{tab:performance}
    Performance comparison (\%).
    Results that are statistically significantly better ($\alpha=0.01$) than \emph{Tree RankGAM} are marked with an asterisk ($^*$), and best GAM result per column is bolded.
  }
\vspace{-0.1in}
{
  \begin{tabular}{|@{}c@{}|c||c|c|c|}
    \hline
        Data set & Method & NDCG$_1$ & NDCG$_5$ & NDCG$_{10}$    \\ \hline \hline 
        \multirow{4}{*}{YAHOO}
            & Tree GAM & 67.61 & 69.46 & 73.89   \\ \cline{2-5}
            & Neural GAM & 67.63 & 69.62 & 73.98   \\ \cline{2-5}
            & Tree RankGAM & 69.12 & 71.03 & 75.04 \\ \cline{2-5}
            & Neural RankGAM & \textbf{69.36} & \textbf{71.32} & \textbf{75.33}$^*$ \\ \hline \hline 
        \multirow{4}{*}{WEB30K}
            & Tree GAM &  29.79 & 32.79 & 35.96   \\ \cline{2-5} 
            & Neural GAM &  30.59 & 33.55 & 36.54   \\ \cline{2-5} 
            & Tree RankGAM & 41.90 & 42.04 & 44.37 \\ \cline{2-5}
            & Neural RankGAM & \textbf{44.31}$^*$ & \textbf{43.29}$^*$ & \textbf{45.09}$^*$  \\ \hline \hline 
        \multirow{5}{*}{CWS}
            & Tree GAM &  19.74 & 32.91 & 36.72   \\ \cline{2-5} 
            & Neural GAM &   20.09 & 34.01 & 38.60   \\ \cline{2-5} 
            & Tree RankGAM & 20.16 & 35.06 & 39.27 \\ \cline{2-5}
            & Neural RankGAM & 20.35 & 34.94 & 38.93  \\ \cline{2-5}
            & Neural RankGAM$+$ & \textbf{24.43}$^*$ & \textbf{39.88}$^*$ & \textbf{42.84}$^*$ \\ \hline 
  \end{tabular}
}
\vspace{-0.2in}
\end{table}

\subsection{Interpretability Analysis (RQ2)}
\label{subsubsec:interpretability}
It is still an ongoing debate~\cite{doshi2017towards,gilpin2018explaining,poursabzi2018manipulating} on how to quantitatively measure model interpretability.
Considering the challenges and the lack of consensus on a rigorous method to evaluate model interpretability, we instead only focus on the family of GAMs.
We argue that our proposed models have similar, if not better, interpretability as traditional GAMs on LTR problems.  

The major differences between our proposed neural ranking GAMs and traditional GAMs are the adoption of ranking losses, the integration of context features, and the instantiation by neural network structures.  We show how these differences affect the model interpretability respectively.

\vpara{RankGAM vs. GAM.}
First, we show that optimizing ranking losses benefits the visualization of sub-model $f_j(\cdot)$ in LTR tasks.
Specifically, we study the the correlation between the effective visual range of $f_j(\cdot)$ curve and the actual importance of the $j$-th feature.
We define the ``\emph{feature importance}'' of the $j$-th feature $x_j$ in a specific model by the NDCG$_5$ drop on validation data sets if perturbing $x_j$ values within every lists by random shuffling. 
We denote this as $\Delta\text{NDCG}_5 (j)$.  
We also define the ``\emph{effective range}'' ($\Delta f_j$) of a sub-model $f_j(\cdot)$ by the difference between its maximum and minimum values evaluated from a sample of $x_j$ values, excluding the bottom and top 5\%.  
Ideally, if the learned sub-models can intuitively reflect the impact of their corresponding features on the final predicted results, we should observe a strong correlation between $\Delta\text{NDCG}_5 (j)$ and $\Delta f_j$.

\begin{figure}[t]
  \centering
  \subfigure[Neural GAM]{
    \label{subfig:web30k_mse_delta_ndcg}
    \includegraphics[width=0.4\columnwidth]{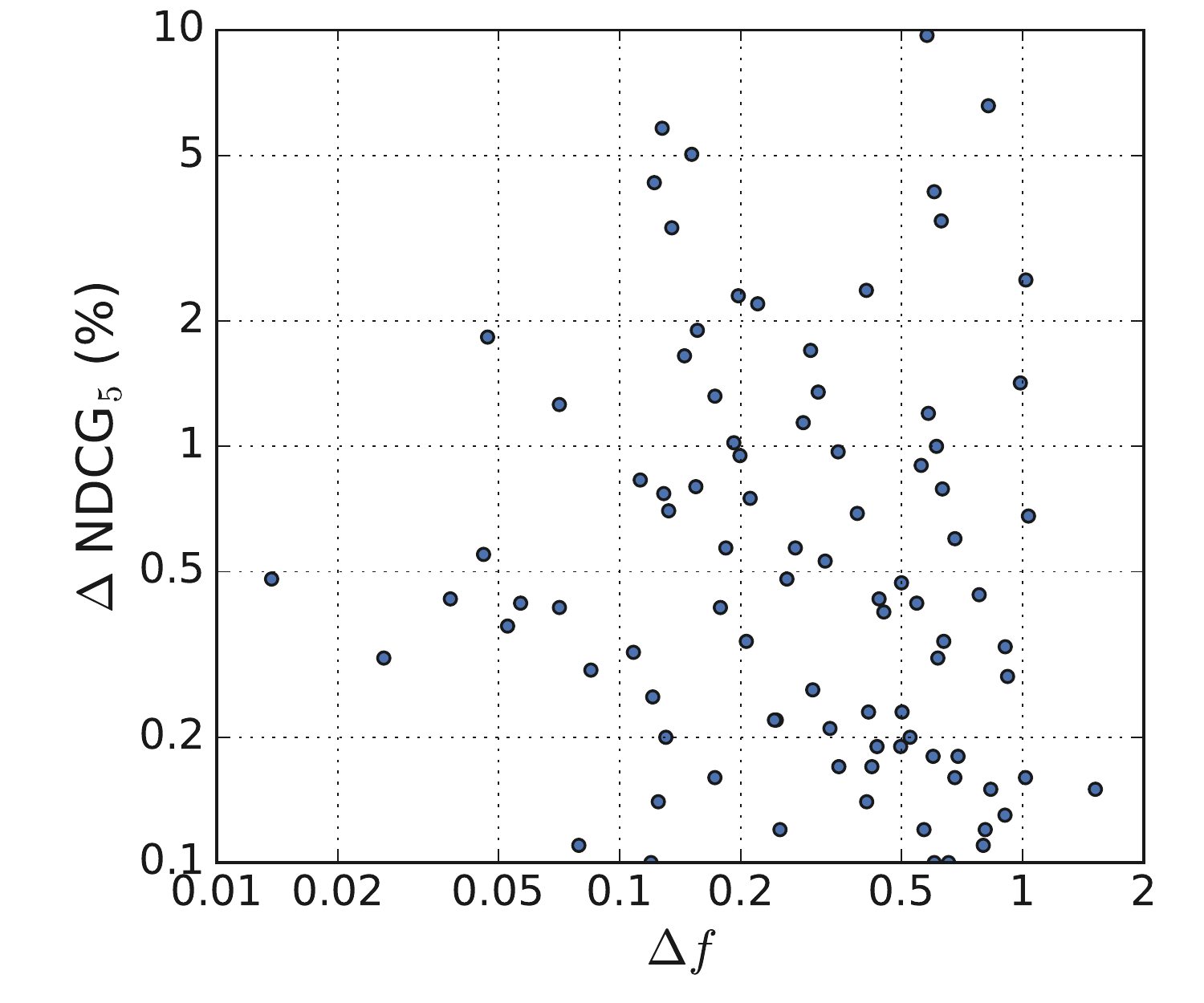}
  }
  \subfigure[Neural RankGAM]{
    \label{subfig:web30k_delta_ndcg}
    \includegraphics[width=0.4\columnwidth]{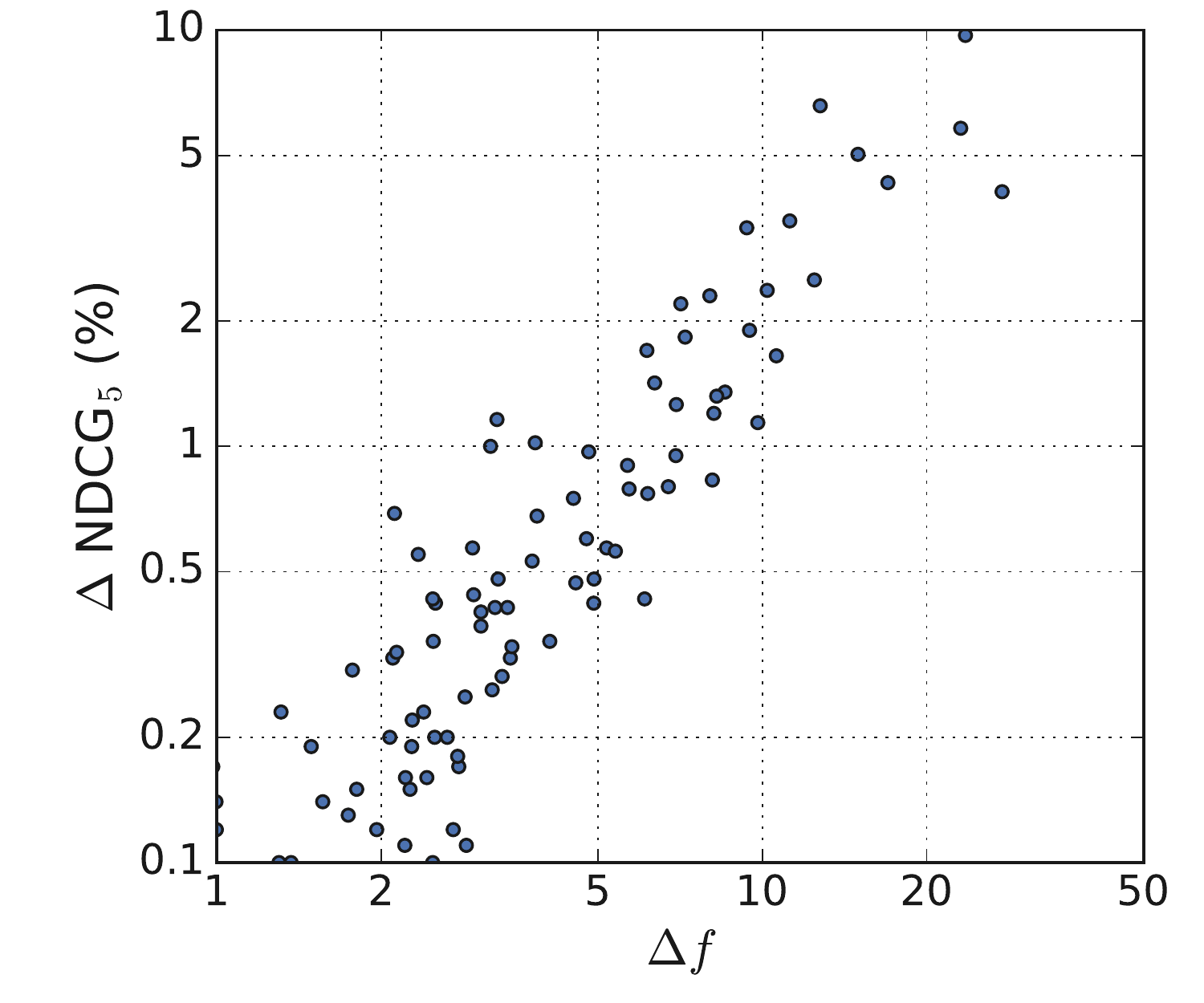}
  }
  \vspace{-0.1in}
  \caption{\label{fig:delta_ndcg}
  Correlation between sub-model effective range $\Delta f$ and its corresponding feature importance $\Delta\text{NDCG}_5$ on WEB30K data set.  Each point corresponds to a feature.  
  }
\end{figure}

To verify this, we plot the $\Delta f_j$ and $\Delta\text{NDCG}_5 (j)$ values for all features on WEB30K data set.  
We conduct the experiments comparing Neural GAM trained with a regression loss (MSE) and Neural RankGAM trained with a ranking loss (ApproxNDCG). 
Figure~\ref{fig:delta_ndcg} shows the results, where each point corresponds to the $\Delta f$ value and the $\Delta\text{NDCG}_5$ value of a feature.  
As one can observe from Figure~\ref{subfig:web30k_delta_ndcg}, the sub-models learned with a ranking loss show a positive correlation between the sub-model's effective range and the importance of the corresponding feature in the final ranking performance.
In comparison, as Figure~\ref{subfig:web30k_mse_delta_ndcg} presents, the sub-models learned with a regression loss does not necessarily reflect their feature importance on ranking performance.  
This is primarily due to the nature of ranking problem where correctly identifying high-ranked items are valued more than distinguishing between massive low-ranked items.
The results again verify the necessity to develop ranking specific GAMs as the interpretability can also be improved. 

\begin{figure}[t]
  \centering
  \subfigure[Feature 7]{
    \label{subfig:tgam_seven_day_click_conversion_rate}
    \includegraphics[width=0.3\columnwidth]{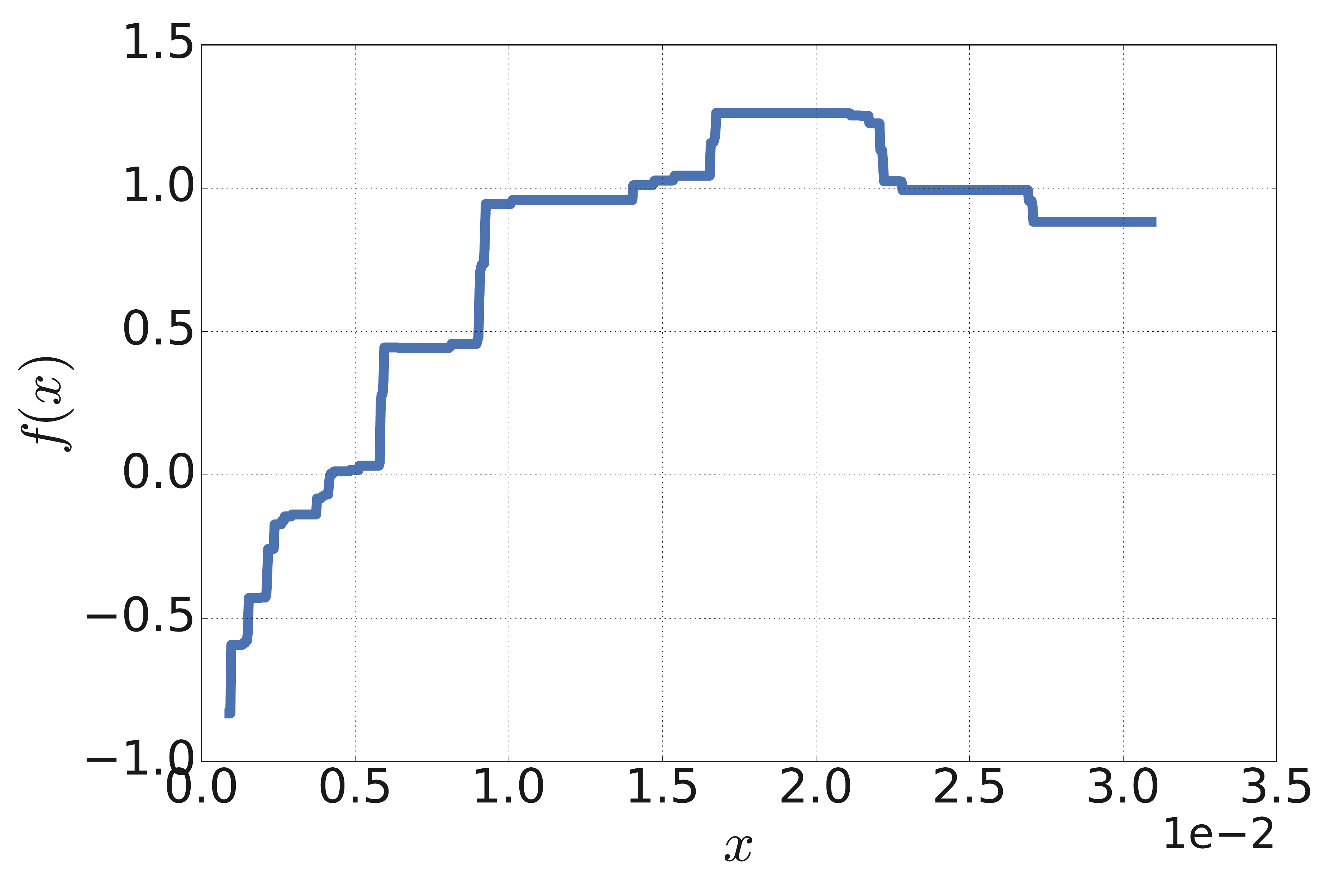}
  }
  \subfigure[Feature 11]{
    \label{subfig:tgam_seven_day_install_conversion_rate}
    \includegraphics[width=0.3\columnwidth]{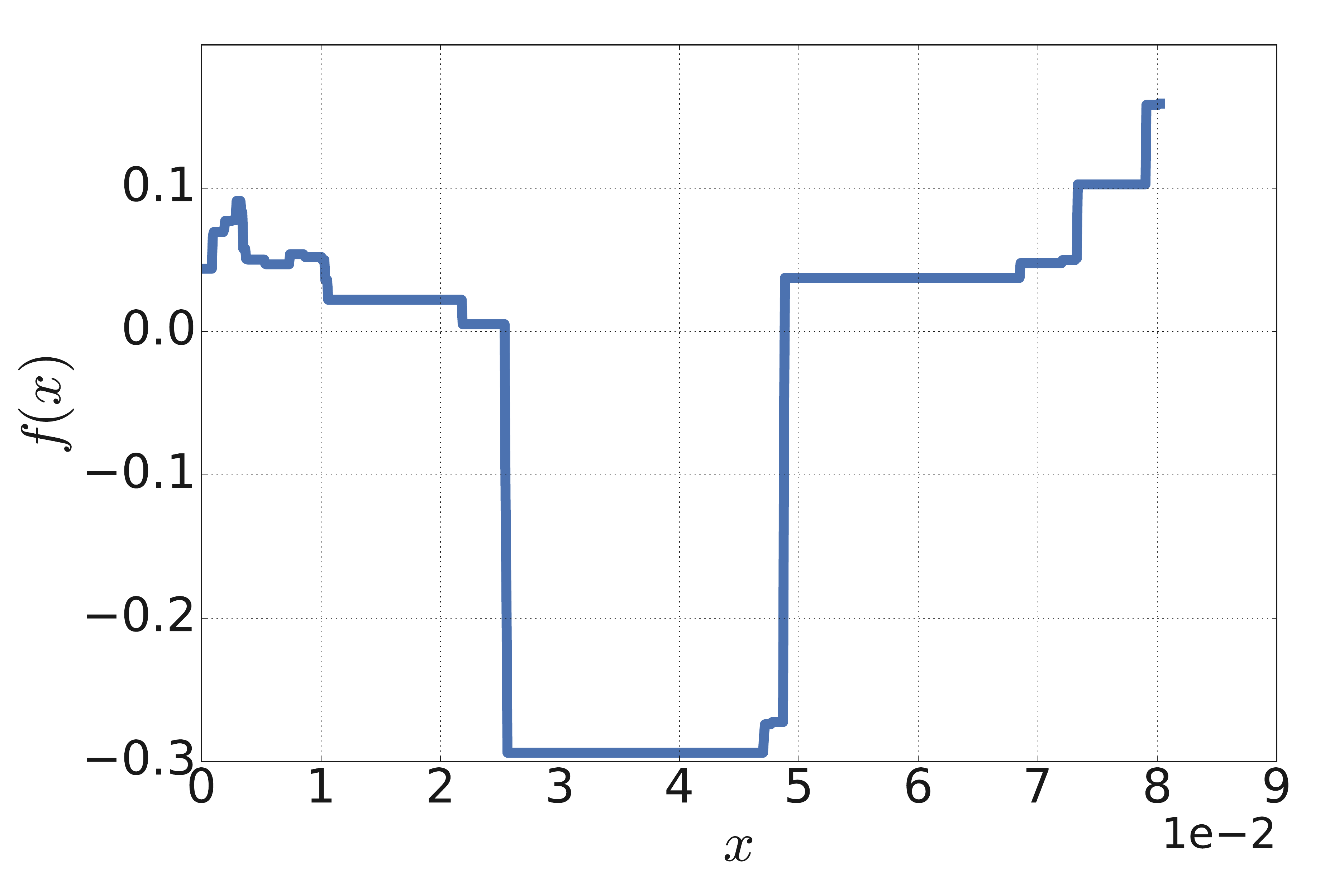}
  }
  \subfigure[Feature 15]{
    \label{subfig:tgam_seven_day_user_flags}
    \includegraphics[width=0.3\columnwidth]{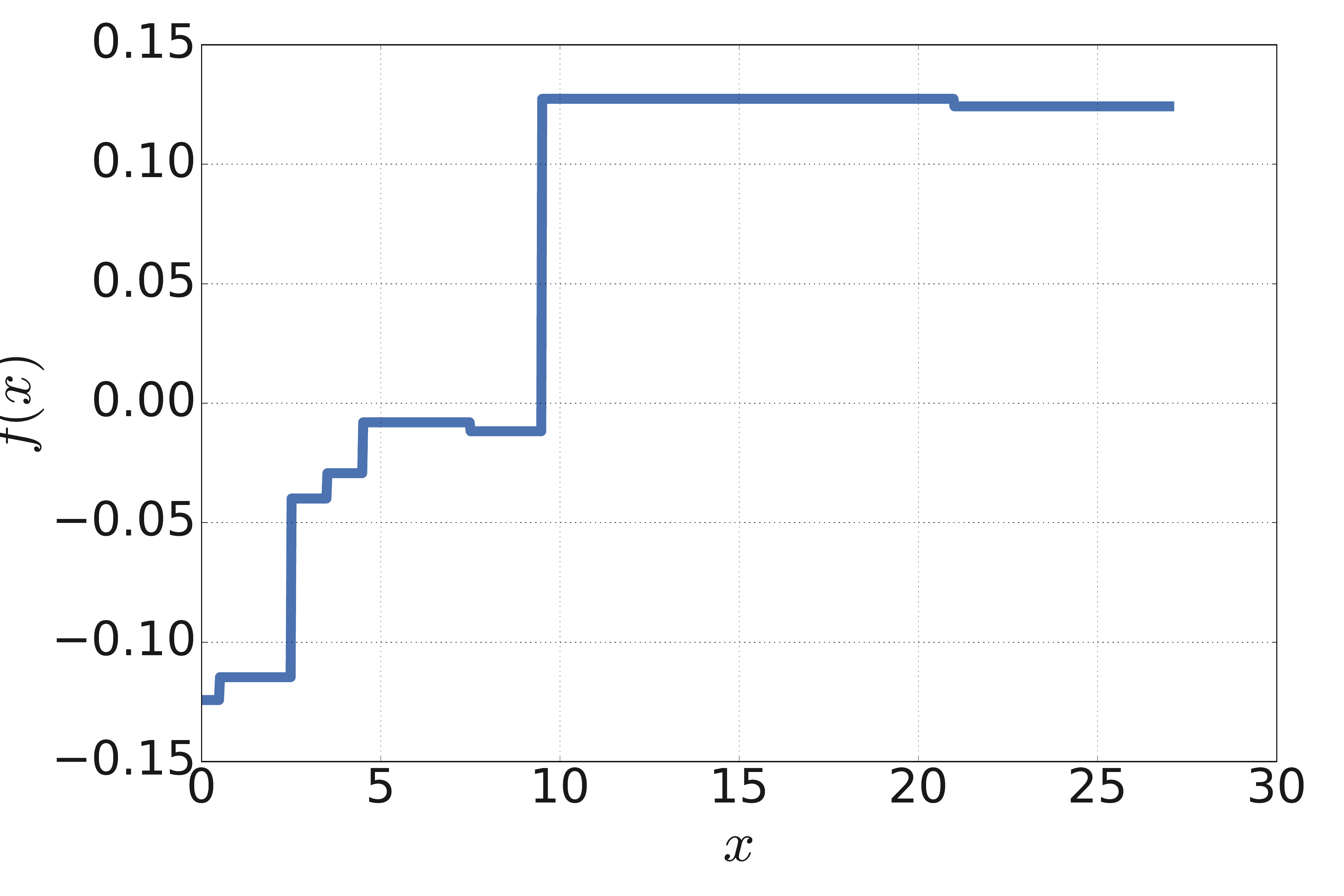}
  }

  \centering
  \subfigure[Feature 7]{
    \label{subfig:ngam_seven_day_click_conversion_rate}
    \includegraphics[width=0.3\columnwidth]{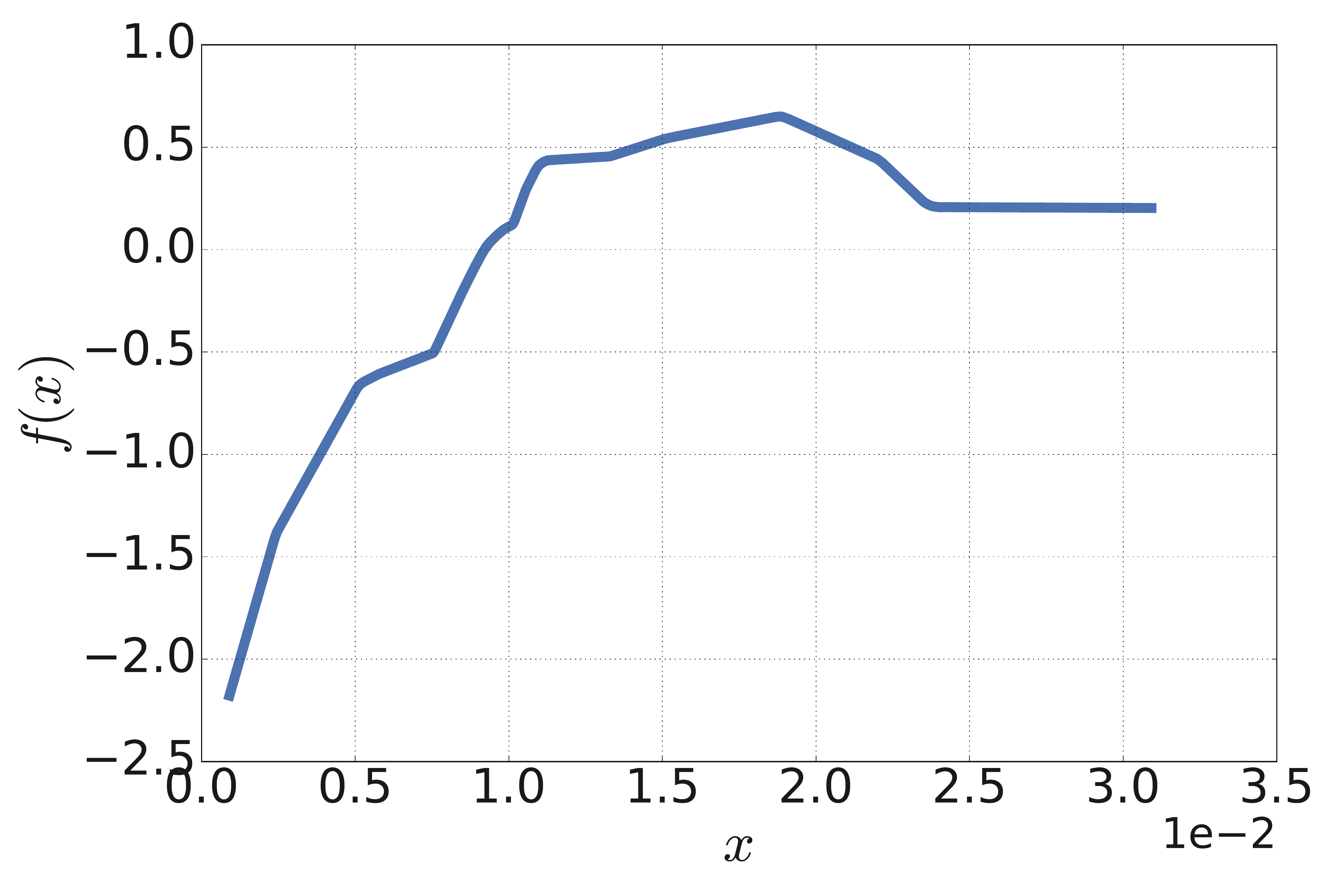}
  }
  \subfigure[Feature 11]{
    \label{subfig:ngam_seven_day_install_conversion_rate}
    \includegraphics[width=0.3\columnwidth]{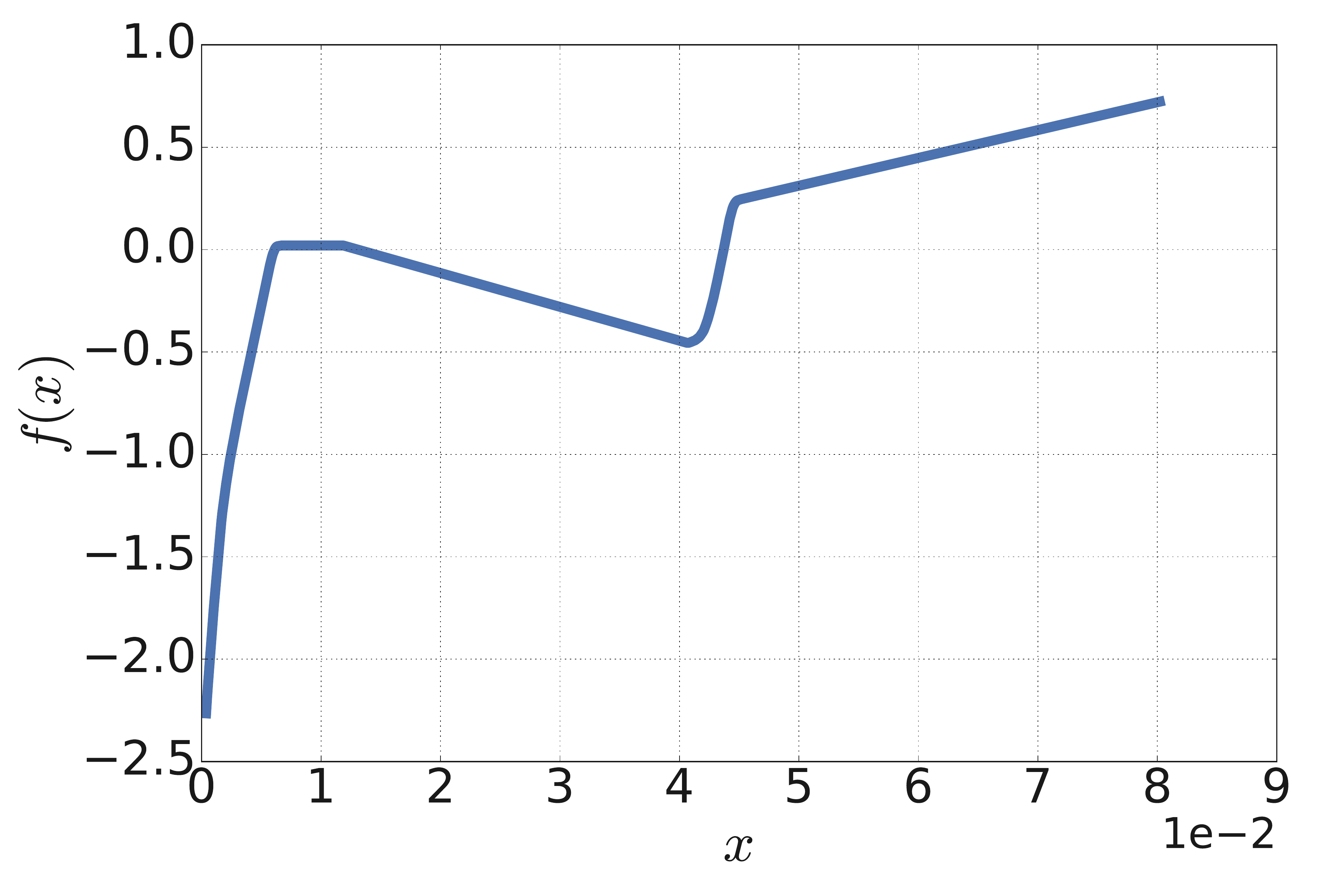}
  }
  \subfigure[Feature 15]{
    \label{subfig:ngam_seven_day_user_flags}
    \includegraphics[width=0.3\columnwidth]{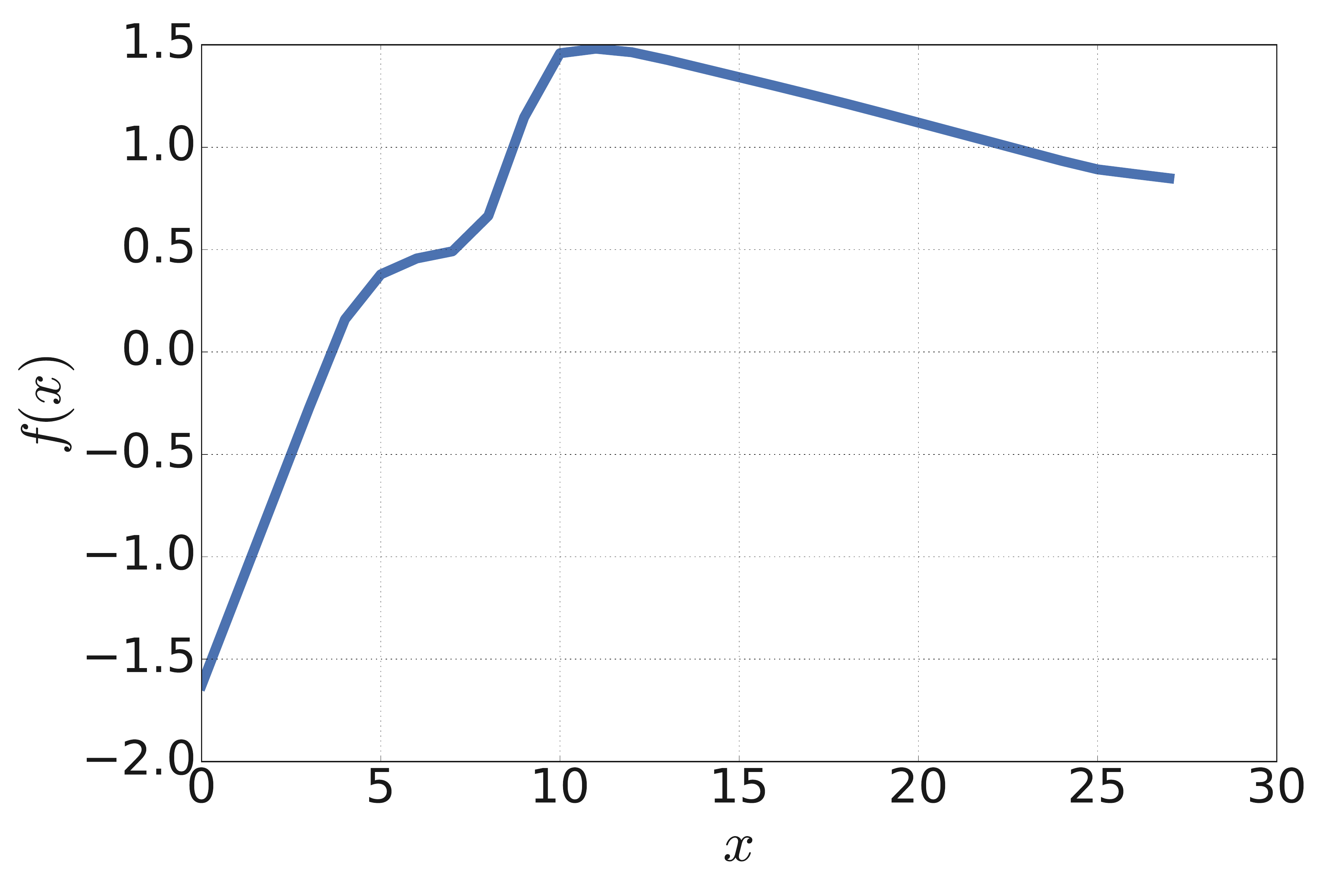}
  }

  \caption{\label{fig:ngam_item_feature_curves}
  Learned sub-model $f(x)$ of Tree RankGAM (top) and Neural RankGAM$+$ (bottom) for selected features on~\dataset~data set.
  }
\end{figure}

\vpara{Neural vs. Tree RankGAM.} 
Second, we compare the interpretability of neural sub-models and tree-based sub-models by visualizing sub-models of the most important features\footnote{The importance of a feature is measured by $\Delta \text{NDCG}_5$ as defined above.} for both Tree RankGAM and Neural RankGAM$+$ on CWS data set.
For each feature, we estimate the distribution and plot the $f(x)$ function for $x$ values above the bottom 5\% and below the top 5\% of all data points in the training data. Figure~\ref{fig:ngam_item_feature_curves} show the curves learned by Tree RankGAM and Neural RankGAM$+$. 

As one can observe, the curves of each feature learned by both methods seem to have similar trends, which confirms the neural sub-models are as effective as tree-based sub-models.
We also notice that the curves in Tree RankGAM are non-continuous, with potentially steep ``steps'' (\eg~around $x=2.5$ in Figure~\ref{subfig:tgam_seven_day_install_conversion_rate} and around $x=5$ in Figure~\ref{subfig:tgam_seven_day_user_flags}), 
whereas the curves learned by Neural RankGAM$+$ with ReLU activation functions are continuous with less steep slopes.  The more continuous curves usually can generalize better, but largely depend on the real applications.  By changing the activation function in Neural RankGAMs, it is also possible to change the corresponding behaviors.

\begin{figure}[t]
\centering
\includegraphics[width=0.9\linewidth]{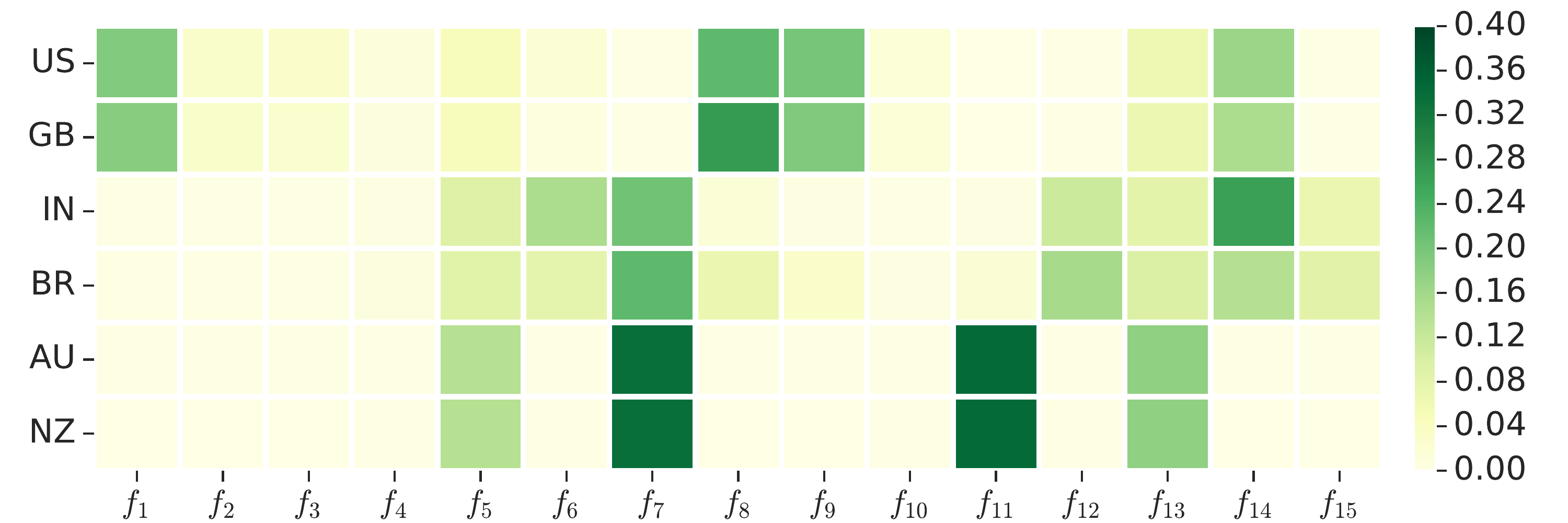}
\caption{
A case study of context feature ``region''.
Each row corresponds to a specific context feature value, where the $j$-th column corresponds to the derived importance weights on item feature sub-model $f_j(x_j)$.  
}
\label{fig:case_region}
\vspace{-0.1in}
\end{figure}

\vpara{Context feature interpretability.} Finally, we show how to visualize the contribution of context features so as to achieve interpretability.
Figure~\ref{fig:case_region} illustrates how item features are weighed based on different values of a specific context feature ``region''.  
This visualization is sufficient for users to discover and analyze patterns of context feature contribution to the final results.
As a running example, it can be observed in Figure~\ref{fig:case_region} that the learned feature importance vectors seem to be similar across regions with high geographical and/or cultural proximity.  
For instance, the US and the GB have similar feature importance vectors as they share similar languages and cultures.

\subsection{Sub-Model Distillation} 
We also show how sub-model distillation can further improve the scalability of our models.
We select Neural RankGAM on YAHOO and WEB30K and Neural RankGAM$+$ on the \dataset~data set. 
We use a 5-segment piece-wise linear functions to distill each sub-model $f_i(\cdot)$ in above selected models.
Table~\ref{tab:distilled_performance} compares the performance before and after the distillation, where the distilled ones are named with a ``(D)'' suffix. The results show that the distilled models have only small NDCG losses.  
On all the data sets, the drop of performance in terms of $\text{NDCG}_{10}$ are less than $-1\%$.
Moreover, as Table~\ref{tab:distilled_model_latency} shows, distilling sub-models substantially reduces model inference time.
On both YAHOO and WEB30K data sets\footnote{We do not disclose the inference times for the \dataset~data set due to its proprietary nature, but the conclusions are similar.}, sub-model distillation achieves 17-23X speed-up.

\begin{figure}[t]
  \centering
  \subfigure[min of term frequency (whole document)]{
    \label{subfig:curve_fitting_1}
    \includegraphics[width=0.3\columnwidth]{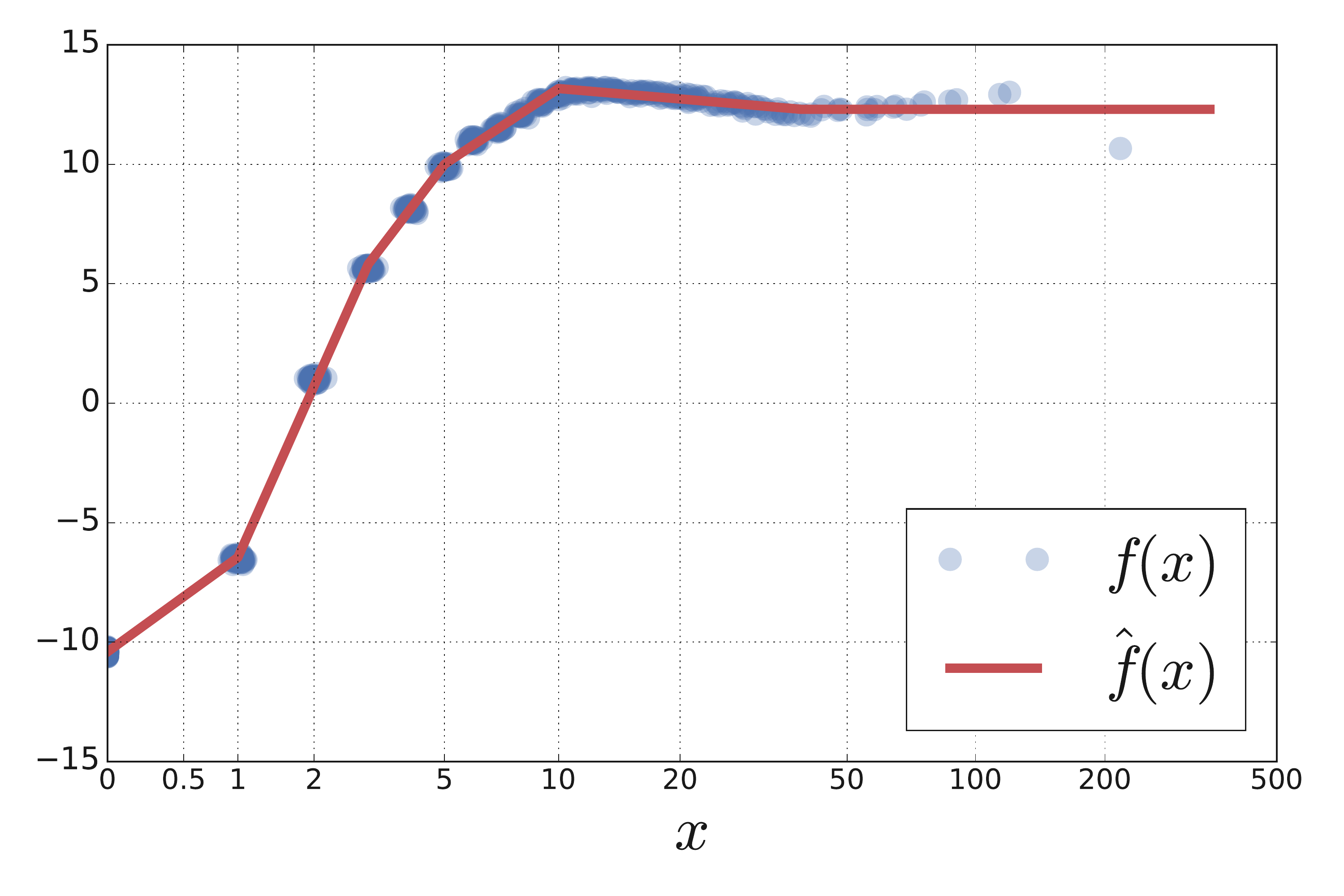}
  }
  \subfigure[LMIR.JM (body)]{
    \label{subfig:curve_fitting_2}
    \includegraphics[width=0.3\columnwidth]{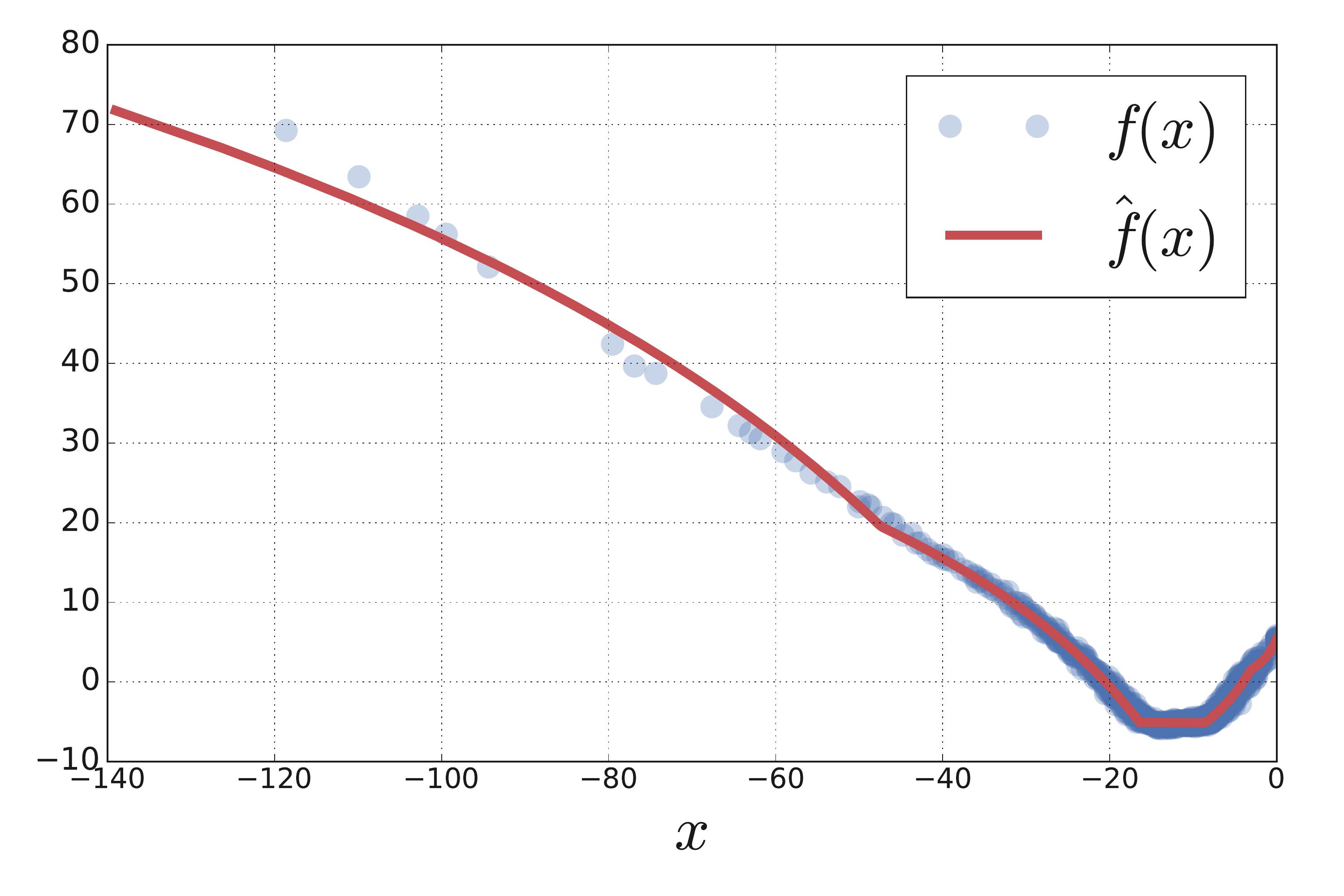}
  }
  \subfigure[sum of stream length normalized term frequency (whole document)]{
    \label{subfig:curve_fitting_3}
    \includegraphics[width=0.3\columnwidth]{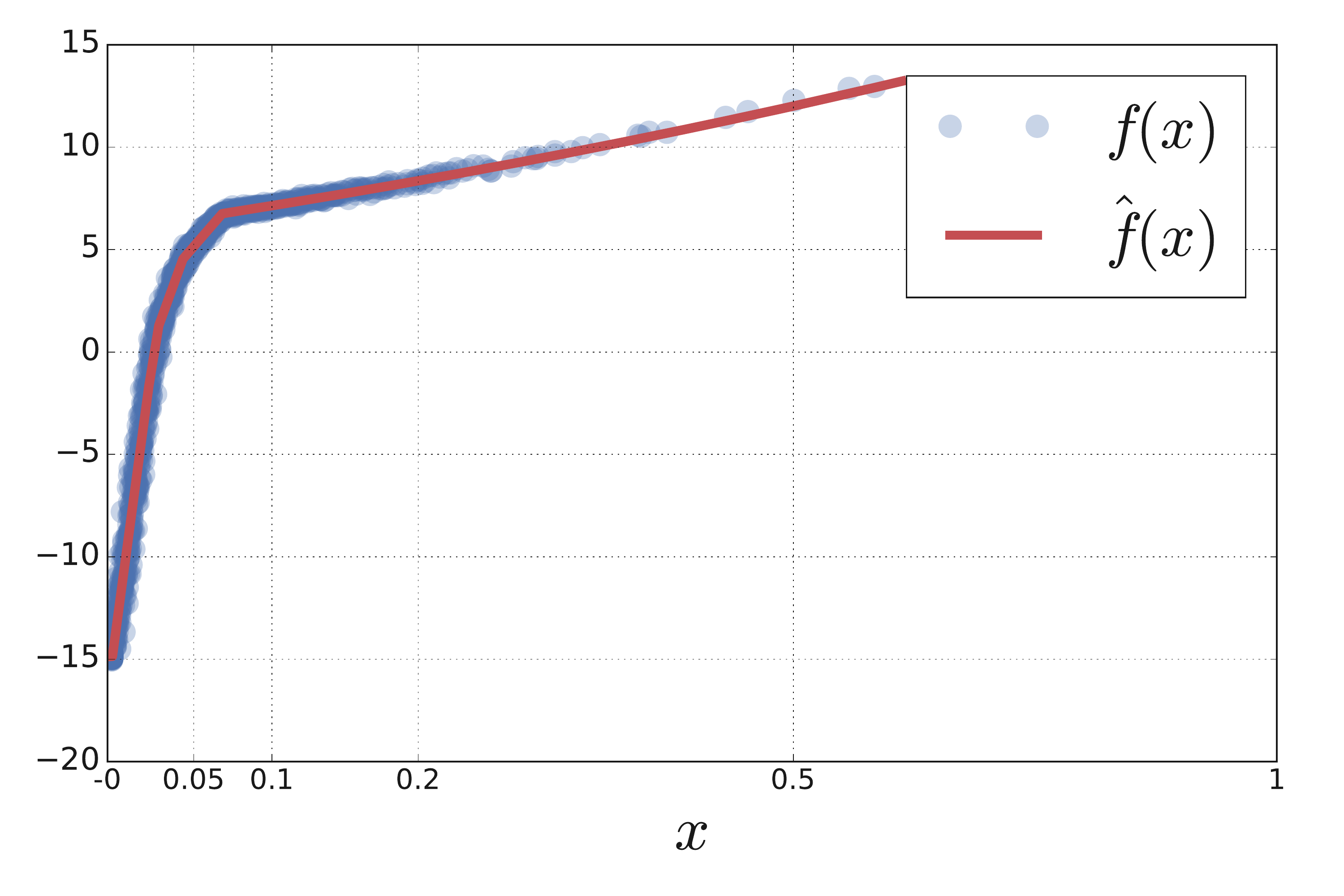}
  }
\vspace{-0.1in}
  \caption{\label{fig:curve_fitting}
  Curves of distilled sub-model $\hat{f}(x)$ and the originally learned sub-model $f(x)$ values in Neural RankGAM on randomly sampled training data points for selected features on WEB30K data set.
  }
\vspace{-0.1in}
\end{figure}

To better visualize the distilled sub-models, we select a set of most important features for the Neural RankGAM on WEB30K data set.
The feature importance is measured by $\Delta\text{NDCG}_{5}$ as defined in Section~\ref{subsubsec:interpretability}.
Figure~\ref{fig:curve_fitting} plots the value of originally learned $f_j(x)$'s at some randomly sampled data points from the training data set and the distilled $\hat{f}_j(x)$'s for the selected features. Clearly, the piece-wise linear function fits the data very well, especially in dense areas with a lot of data points from training data.  

Admittedly, it could be observed that the error of the distilled sub-model in sparse areas with fewer data points from training data could be larger (\eg~around $x=200$ in Figure~\ref{subfig:curve_fitting_1} or around $x=-120$ in Figure~\ref{subfig:curve_fitting_2}).  
Data points around these areas may actually have high ranking scores.
Hence, errors for these data points might lead to larger performance drops in terms of ranking metrics, as shown in the relatively larger performance drop on WEB30K data set.
Developing a sub-model distillation technique specifically for ranking is another non-trivial task.
Therefore, we leave it for future study.

\begin{table}[t]
\centering
  \caption{
  \label{tab:distilled_performance}
    Comparing performances before and after sub-model distillation.
    Models with distilled sub-models is marked with suffix ``(D)''.
  }
  \vspace{-0.1in}
{
  \begin{tabular}{|@{}c@{}|@{}c@{}||c|c|c|}
    \hline
        Data set & Method & NDCG$_1$ & NDCG$_5$ & NDCG$_{10}$    \\ \hline \hline 
        \multirow{2}{*}{YAHOO}
            & Neural RankGAM &  69.36 & 71.32 & 75.33    \\ \cline{2-5}
            & Neural RankGAM (D) & 69.41 & 71.33 & 75.33   \\ \hline \hline 
        \multirow{2}{*}{WEB30K}
            & Neural RankGAM &  44.31 & 43.29 & 45.09   \\ \cline{2-5} 
            & Neural RankGAM (D) & 41.63 & 41.90 & 44.15 \\ \hline \hline
        \multirow{2}{*}{\dataset}
            & Neural RankGAM$+$ & 24.43 & 39.88 & 42.84    \\ \cline{2-5} 
            & Neural RankGAM$+$ (D) & 24.11 & 39.42 & 42.39 \\ \hline
  \end{tabular}
}
\end{table}

\begin{table}[t]
\centering
  \caption{
  \label{tab:distilled_model_latency}
    Comparison of model inference time $\tau$ before and after sub-model distillation measured by CPU time.
  }
  \vspace{-0.1in}
{
\begin{tabular}{|@{}c@{}||c|c||c|c|}
\hline
Data set  & \multicolumn{2}{c||}{YAHOO} & \multicolumn{2}{c|}{WEB30K}    \\ \hline
Method &  Total $\tau$ & $\tau$ / query   &  Total $\tau$ & $\tau$ / query  \\ \hline \hline
Neural RankGAM   & 327s & 47ms & 303s & 48ms \\ \hline
Neural RankGAM (D) & 19s & 3ms & 13s & 2ms \\ \hline
\end{tabular}
}
\vspace{-0.1in}
\end{table}

\begin{figure}[t]
  \centering
  \subfigure[YAHOO]{
    \label{subfig:yahoo_hidden_layer}
    \includegraphics[width=0.4\columnwidth]{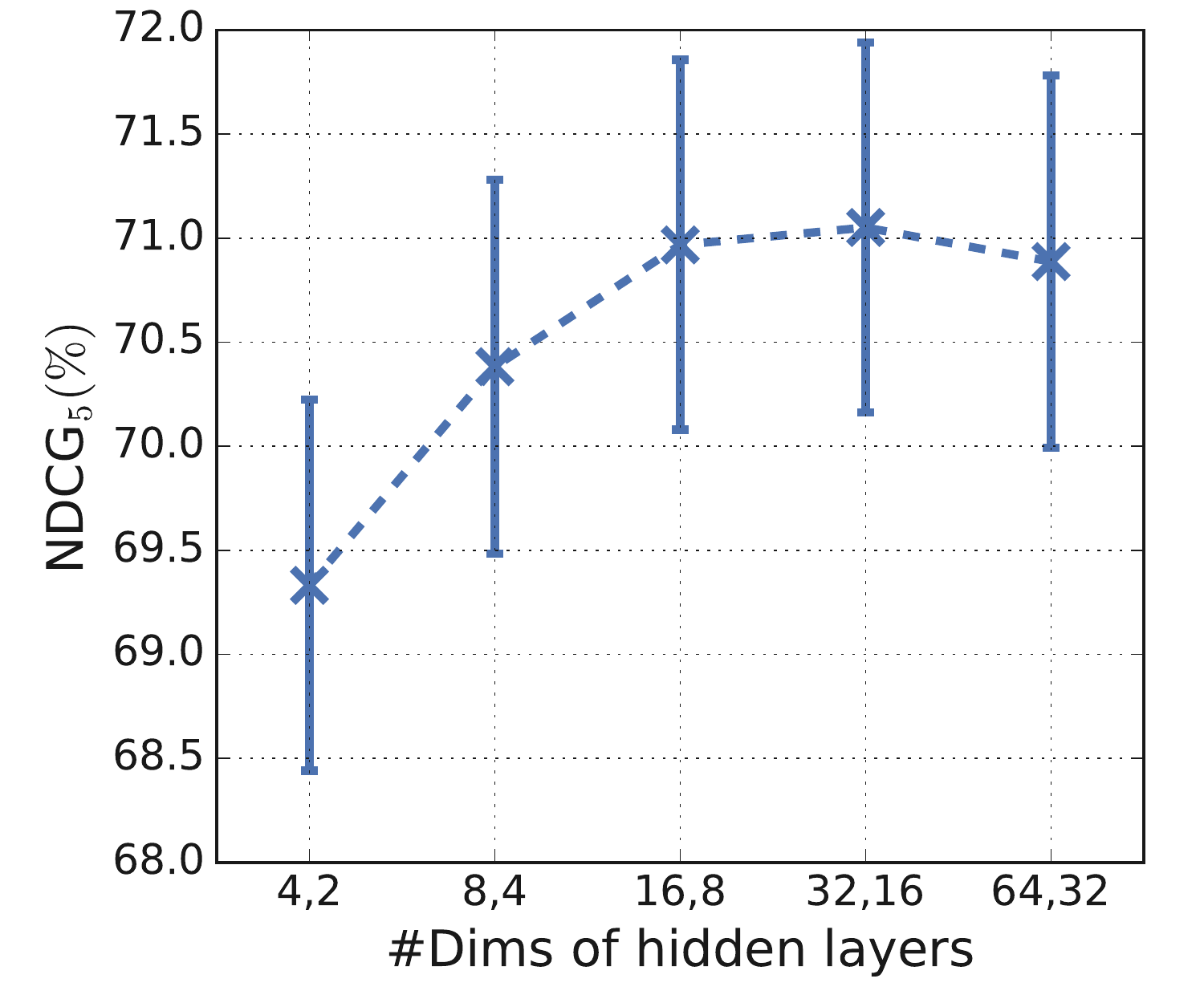}
  }
  \subfigure[WEB30K]{
    \label{subfig:web30k_hidden_layer}
    \includegraphics[width=0.4\columnwidth]{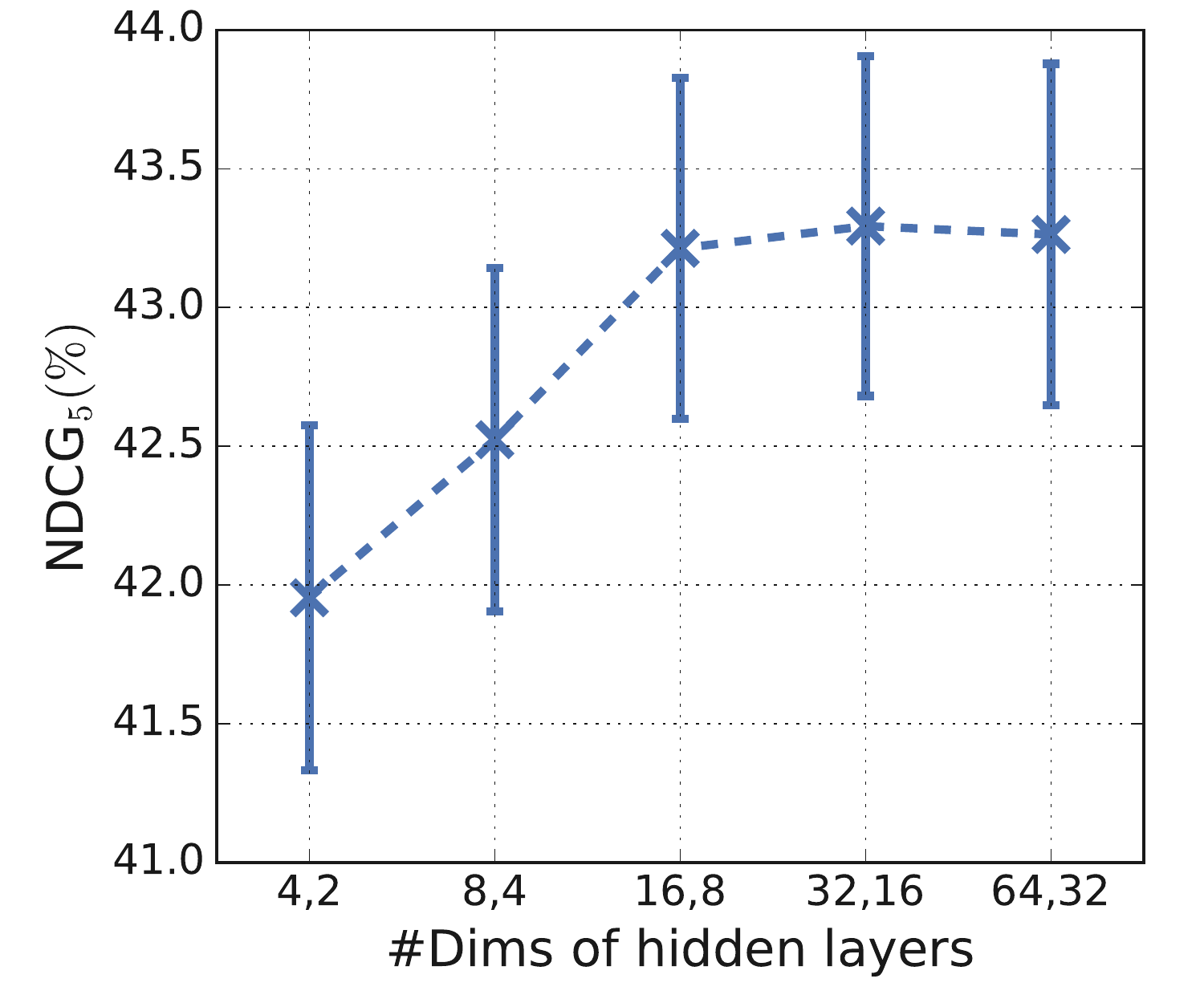}
  }
  \vspace{-0.1in}
  \caption{\label{fig:hidden_layer}
  Comparing performances of Neural RankGAM with different number of dimensions in the hidden layers (95\% confidence interval). 
  }
  \vspace{-0.1in}
\end{figure}

Each sub-model does not need to be a high capacity model. In Figure~\ref{fig:hidden_layer},
we fix the number of hidden layers $L=2$ but vary the number of dimensions for hidden layers from $(4,2)$, $(8,4)$ to $(64,32)$ on both the YAHOO and WEB30K data sets. 
The performance basically saturates for hidden layer dimensions greater than $(16,8)$. It verifies that a relative simple network is sufficient for sub-models.


\begin{table}[]
\centering
  \caption{
  \label{tab:stacked_dnn}
  Performance comparison of Neural RankGAM, a black-box DNN and a hybrid model combining both on WEB30K.
  Best results are bolded.  Results significantly better ($\alpha=0.05$) than DNN are marked with asterisks ($^*$).
  }
  \vspace{-0.1in}
{
  \begin{tabular}{|c||c|c|c|}
    \hline
        Method & NDCG$_1$ & NDCG$_5$ & NDCG$_{10}$    \\ \hline \hline 
            Neural RankGAM & 44.31 & 43.29 & 45.09     \\ \hline
            DNN~\cite{bruch2019SIGIR} &  46.64 & 45.38 & 47.31    \\ \hline
            Hybrid &  \textbf{47.54} & \textbf{46.26}$^*$ & \textbf{48.21}$^*$  \\ \hline 
  \end{tabular}
}
\vspace{-0.1in}
\end{table}





\subsection{Improving Black-Box Models} 
We show that on some data sets, striving for interpretability can also provide insights for improving black-box DNN models.
Specifically, we take the trained Neural RankGAM and use sub-model outputs as features to feed into a fully-connected feed-forward network.
The fully-connected neural networks have similar configuration as~\cite{bruch2019SIGIR}.
Table~\ref{tab:stacked_dnn} shows our preliminary results on WEB30K.  
It can be shown that this hybrid model can further improve the performance of the vanilla DNN model, whereas simply increasing the number of layers or neurons does not~\cite{bruch2019SIGIR}.
This might provide some insights on how training GAM-style models can effectively learn some knowledge from the data that can eventually benefit all other models.

\section{Conclusions}
\label{sec:conclusion}

In this paper, we explore building \emph{intrinsically interpretable} learning-to-rank models by adapting generalized additive models (GAMs). 
Different from previous GAMs that mainly focus on regression or classification, 
our proposed ranking GAMs can capture both list-level context features and item-level features of ranking tasks. 
While traditional GAMs are instantiated with splines or trees, we choose neural networks for more flexible model structures, training objectives and feature types.
Our experiments verify that the proposed ranking GAMs outperform traditional GAMs on LTR tasks and effectively leverage list-level context features to further improve the performance, while maintaining interpretability of GAMs. 
We also present how to distill sub-models into simple piece-wise linear functions for further simplicity.

As an early exploration on building intrinsically interpretable LTR models, we stick to the most basic form of GAM.
However, we are aware that there are numerous directions to further develop and understand ranking GAMs:
(1) Incorporating limited pairwise feature interactions to improve model accuracy with reasonable interpretability (similar to~\cite{lou2013accurate});
(2) Enabling users to dictate constraints such as monotonicity on sub-models;
(3) Combining ranking GAMs with fully-fledged neural networks or decision tree models to achieve both high accuracy and high interpretability;
(4) Studying the formal definition and the evaluation methodology of ranking model interpretability.




\appendix
\section*{Acknowledgments}

We thank Vytenis Sakenas, Dmitry Osmakov, Olexiy Oryeshko for implementing early version of Neural RankGAM and piece-wise regression for us. We also thank Po Hu, Janelle Lee, and Chary Chen for preparing data sets for our experiments.


\bibliographystyle{ACM-Reference-Format}
\bibliography{references}

\end{document}